\documentclass[referee]{aa}

\usepackage{pgf}
\usepackage{graphicx}
\usepackage{txfonts}
\usepackage{natbib,twoopt}
\usepackage{color}

\bibpunct{(}{)}{;}{a}{}{,}             
\makeatletter
  \newcommandtwoopt{\citeads}[3][][]{\href{http://adsabs.harvard.edu/abs/#3}%
    {\def\hyper@linkstart##1##2{}%
     \let\hyper@linkend\@empty\citealp[#1][#2]{#3}}}
  \newcommandtwoopt{\citepads}[3][][]{\href{http://adsabs.harvard.edu/abs/#3}%
    {\def\hyper@linkstart##1##2{}%
    \let\hyper@linkend\@empty\citep[#1][#2]{#3}}}
  \newcommandtwoopt{\citetads}[3][][]{\href{http://adsabs.harvard.edu/abs/#3}%
   {\def\hyper@linkstart##1##2{}%
     \let\hyper@linkend\@empty\citet[#1][#2]{#3}}}
  \newcommandtwoopt{\citeyearads}[3][][]%
    {\href{http://adsabs.harvard.edu/abs/#3}
    {\def\hyper@linkstart##1##2{}%
     \let\hyper@linkend\@empty\citeyear[#1][#2]{#3}}}
 \makeatother

\def\kms{$\rm km\;s^{-1}$}

\def\gtsim{\mathrel{\spose{\lower.5ex \hbox{$\mathchar"218$}}
     \raise.4ex\hbox{$\mathchar"13E$}}}
\def\ltsim{\mathrel{\spose{\lower.5ex\hbox{$\mathchar"218$}}
     \raise.4ex\hbox{$\mathchar"13C$}}}
\def\aFe{[$\alpha/{\rm Fe}$]}

\def\Hb{${\rm H}{\beta}$}
\def\hb{${\rm H}{\beta}$}

\def\Mgb{{\rm Mg}\,$_{\rm b}$}
\def\Fe{$\langle {\rm Fe}\rangle$}

\def\ZH{[$Z/{\rm H}$]}
\def\MgFe{[${\rm MgFe}$]$'$}

\def\Mgd{{\rm Mg}$_{\rm 2}$}

\def\kms{$\rm km\,s^{-1}$}

\def\apj{ApJ}
\def\aj{AJ}
\def\apjl{ApJL}
\def\apjs{ApJS}
\def\mnras{MNRAS}
\def\aaps{A\&AS}
\def\aap{A\&A}

\def\pasp{PASP}

\def\spose#1{\hbox to 0pt{#1\hss}}

\def\aj{AJ}                   
\def\apj{ApJ}                 
\def\apjl{ApJ}                
\def\apjs{ApJS}               
\def\aap{A\&A}                
\def\aaps{A\&AS}              
\def\mnras{MNRAS}             
\def\pasp{PASP}               

\def\procspie{Proc.~SPIE}   

\def\kms{$\rm km\,s^{-1}$}

\def\Hb{${\rm H}{\small{\beta}}$}

\begin{document}

\title{Near Infrared spectroscopic indices for unresolved stellar populations. I. 
Template galaxy spectra.\thanks{Based on observations made with ESO Telescopes 
at the La Silla Paranal Observatory under programme ID 086.B-0900(A)}}

\author{
P. Fran\c{c}ois\inst{1}
\and
L. Morelli\inst{2,3,4}
\and
A. Pizzella\inst{3,4}
\and
V. D. Ivanov\inst{5,6}
\and
L. Coccato\inst{5}
\and
M. Cesetti\inst{3}
\and
E. M. Corsini\inst{3,4}
\and
E. Dalla Bont\`a\inst{3,4}
}

\offprints{Patrick Fran\c {c}ois, \email{patrick.francois@obspm.fr}}

\institute{
GEPI, Observatoire de Paris, PSL Research University, CNRS, Univ. Paris Diderot, 
Sorbonne Paris Cité, 61 avenue de l'Observatoire, 75014, Paris, France 
\and
Instituto de Astronom\`{i}a y Ciencias Planetarias Universidad de Atacama, Copiap\`{o}, Chile
\and
Dipartimento di Fisica e Astronomia "G. Galilei'', Universita di Padova, 
vicolo dell'Osservatorio 3, I-35122 Padova, Italy
\and
INAF-Osservatorio Astronomico di Padova, vicolo dell'Osservatorio 5, I-35122 
Padova, Italy
\and
European Southern Observatory, Karl-Schwarzschild-Str. 2, 
85748 Garching bei M\"unchen, Germany
\and
European Southern Observatory, Ave. Alonso de C\'ordova 3107, 
Vitacura, Santiago, Chile
}

\date{Received  / Accepted }

\abstract 
{A new generation of spectral synthesis models has been developed 
in the recent years, but there is no matching -- in terms of quality 
and resolution -- set of template galaxy spectra for testing and 
refining the new models.}
{Our main goal is to find and calibrate new near-infrared spectral
  indices along the Hubble sequence of galaxies which will be used to
  obtain additional constraints to the population analysis based on
  medium resolution integrated spectra of galaxies.}
{Spectra of previously studied and well understood galaxies with 
relatively simple stellar populations (e.g., ellipticals or bulge 
dominated galaxies) are needed to provide a baseline data set for 
spectral synthesis models.}
{X-Shooter spectra spanning the optical and infrared wavelength
  (350-2400\,nm) of bright nearby elliptical galaxies with resolving
  power $ R\sim$4000-5400 were obtained. Heliocentric systemic
  velocity, velocity dispersion and Mg, Fe and \Hb\ line-strength indices
  are presented.}
{We present a library of very high quality spectra of
  galaxies covering a large range of age, metallicity and
  morphological type. Such as a dataset of spectra will be crucial to
  address important questions of the modern investigation concerning
  galaxy formation and evolution.}

\authorrunning{P. Fran\c{c}ois et al.}
\titlerunning{Medium resolution spectral library}

\maketitle

\section{Introduction}\label{sec:intro}
Recent advances in the infrared (IR) instruments  like X-Shooter  \citet{2011A&A...536A.105V} and KMOS
 \citep{2013Msngr.151...21S} and their availability at the large telescopes made it possible 
to extend the stellar population studies of unresolved galaxies into the 
IR domain 
\citep[see for example the pioneering work of][]{1980ApJ...238...24R,1988ApJ...325..679R}.
Reduction of the extinction is an obvious, but not the only advantage of 
the expansion towards longer wavelengths -- a look at the color-magnitude 
diagram of any complex stellar system suggests that different stellar 
populations dominate different spectral ranges
\citep[see the discussion in ][]{2011MNRAS.411.1897R,2015ApJS..217...13M}, 
opening up the possibility to disentangle their contribution in these 
systems.

Comprehensive stellar libraries are the first step toward developing the 
necessary tools for IR studies. An incomplete list of the previous work 
includes \citet{1970AJ.....75..785J}, \citet{1986ApJS...62..501K}, 
\citet{1986ApJS...62..373L}, \citet{1993A&A...280..536O}, 
\citet{1997ApJS..111..445W}, \citet{1998AJ....116.2520J},
\citet{1998ApJ...508..397M}, \citet{2000ApJ...535..325W}, 
\citet{2002AJ....124.3393W}, \citet{2004ApJS..151..387I},
\citet{2015AJ....149..181Z}; for more historic references see Table\,1 in
\citet{2009ApJS..185..289R}.
The early theoretical attempts to model the IR stellar spectra faced 
difficulties with the incorporation of millions of molecular lines and 
generated synthetic spectra that did not match well the broad band 
features, because of the complications with the broad molecular features 
\citep{1997A&A...318..841C}. Late type stars with prominent molecular bands 
 were sometimes omitted, even for optical 
spectral models \citep{1992A&A...264..557B}.

These observational libraries quickly found other applications  to derive 
stellar diagnostics in heavily obscured regions 
\citep{1996ApJ...456..206T,1997ApJ...489..698H,2002ApJS..138...35H,2009ApJ...697..701M} 
and for objects that are intrinsically bright in the IR because of their 
low effective temperatures, like brown dwarfs, 
\citep{1995Sci...270.1478O,1996ApJ...467L.101G,2003ApJ...593.1074G,2014A&A...562A.127B}
AGB, red giant and red supergiant stars
\citep{1986ApJS...62..373L,1991ApJ...378..742T,1997AJ....113.1411R,2000A&AS..146..217L,2013ApJ...767....3D}.

A major step in the development of libraries came with the work of 
\citet[a.k.a. IRTF spectral library]{2009ApJS..185..289R}.

 It included 210 
stars, observed with at a resolving power $R\sim$2000, and had a spectral 
coverage from $\lambda$$\sim$0.8\,$\mu$m to 2.5\,$\mu$m. An earlier precursor
paper \citep{2005ApJ...623.1115C} included spectra of ultracool M, L and T-type 
dwarfs observed with the same instrument set up. This is 
the first flux calibrated IR stellar library which facilitated its use for 
modeling the integral spectra of galaxies.

The first generation of evolutionary models with a proper treatment of the 
IR wavelength range \citep{2005MNRAS.362..799M} used theoretical stellar 
spectra \citep{1997A&AS..125..229L,1998A&AS..130...65L} but the availability 
of the empirical IRTF spectral library prompted the development of a number 
of models based on it. \citet{2012ApJ...747...69C} pointed at the presence of 
features, sensitive to the initial mass function  (IMF) slope, age and  \aFe  abundance ratio. 
\citet{2015A&A...582A..96M} carried out extensive preparation of the IRTF 
library and reported in a companion paper \citep{2015A&A...582A..97M} a new 
synthetic model that fitted well the published indices of elliptical galaxies. 
\cite{2015MNRAS.449.2853R} from the same research group extended the 
population models from $\lambda$$\sim$2.5\,$\mu$m to 5\,$\mu$m with the subset 
of the IRTF library spectra that covered this wavelength range. These models 
agree well with the observations for older galaxies, but have problems with 
younger and/or metal-poor populations. Finally, a combined model spanning the 
entire optical-IR range was reported in \citet{2016A&A...589A..73R}, together 
with new tests that further validate its results.

This progress revealed an unexpected problem -- the lack of matching 
high-quality galaxy spectra. An example of a forced ``work around'' resorting 
to broad band photometry for population synthesis purposes is the work of
\citet{2005A&A...443..735C}, but there are also on-going efforts to obtain IR 
spectra of galaxies. Not surprisingly, the first of them were aimed at the 
classes of obscured objects for which these challenging observations yielded 
the biggest advantages: star forming galaxies
\citep{1988ApJ...325..679R,1994ApJ...421..101D,1994ApJ...421..115D,1998ApJ...505..639E,1996ApJ...466..150V,1997ApJ...479..694V,2001A&A...366..439K},
luminous and ultra-luminous IR galaxies
\citep[LIRG and ULIRG;][]{1994ApJ...421..101D,1997ApJ...484...92V,1999ApJ...522..139V}
and galaxies with active nuclei 
\citep[AGN;][]{1997ApJ...477..631V,2000ApJ...545..190I,2002MNRAS.331..154R,2003MNRAS.343..192R,2004ApJ...614..122I,2006A&A...457...61R,2009A&A...499..417H,2012ApJS..203...14M}.
The influx of more and better spectra of galaxies with various types of 
activity continues, but most of these studies lack the signal-to-noise (S/N) to 
investigate the galaxy continuum and  absorption features; instead, they 
concentrated on the emission lines and tried to address general questions 
like what was nature of the hidden (in the optical wavelengths, at least) 
power sources, what was 
the contribution of the AGN -- far from the determining parameters of the 
stellar populations directly from absorption features, as it has been done 
in the optical for years \citep[e.g.,][]{1994ApJS...95..107W}.

Encouragingly, \citet{2015ApJS..217...13M} found some differences in the 
strength of absorption features and the continuum shapes in galaxies with 
younger stellar populations -- e.g., Balmer absorption lines are associated 
with population younger than 1.5\,Gyr. They found the behaviour of some 
metal features promising as well, but their sample was dominated by LINERS 
and Seyferts and only in a few cases the activity was sufficiently low that 
it did not affect significantly the stellar continuum.

The contribution of the AGB stars towards to total galaxy spectra is another 
open issue. Molecular CN and VO bands that originate in thermally pulsating 
  AGB (TP-AGB) stars were also detected by \citet{2015ApJS..217...13M}, but these 
features were weaker in $\sim$1\,Gyr galaxies than in older ones. This may 
be a bias related to the intrinsic velocity dispersion and the 
contamination from other features. Previously \citet{2013MNRAS.428.1479Z} 
had found negligible contribution of TP-AGB stars in post-starburst 
galaxies with ages 0.5-1.5\,Gyr. More recently \citet{2015MNRAS.450.3069R} 
called for a revision of models of \citet{2011MNRAS.418.2785M} to reduce 
the contribution TP-AGB stars. On the other hand \citet{2009ApJ...705L.199M} 
measured systematically stronger $K$-band Na{\sc i} and CO in galaxies with 
AGB population, than in those without it, so the question if younger 
populations can be reliably detected with IR spectroscopy remains open.

 At least one of our galaxies, NGC 7424 has an age estimate (Table \ref{tab:central_values})
that suggests it may contain significant populations of AGB stars.
The age estimates are luminosity weighted and the high intrinsic
luminosity of the AGB stars in the IR range implies that these stars
contribute a notable fraction to the total IR light of the galaxies
(e.g. $\sim$ 17 \% at H band; 
\citet{2012ApJ...748...47M}).

The AGB stars are surrounded by dust envelopes that reprocess the
stellar light and re-emits it to the near- or the mid-IR, although
to obtain a good fit to the observed near-/mid-IR spectral energy
distribution of the galaxies may require to adjust the AGB lifetimes
(e.g. \citet{2015ApJ...806...82V}). The quick and short lived
AGB phase lead to luminosity fluctuations that affects the M/L ratio
\citep{2000ASPC..211...34L,2011ASPC..445..391M}, especially in the IR, and
the star formation rate estimates. The AGB stars are responsible
for the dust production and therefore - the the intrinsic reddening
inside galaxies. Furthermore, the intrinsically red AGB stars modify
the apparent galaxy colors to appear redder at younger age; omitting
the AGB contribution leads to incorrect age and mass estimates
\citep{2006ApJ...652...85M}.

All these effects grow increase with redshit, when progressively
younger galaxies are observed. Therefore, better stellar population
models that include the effects of the AGB stars will be needed to
interpret the observations for the James Webb space telescope, and
are being prepared (e.g. Marigo et al. 2017) and our library is
intended to help testing and validating them.

Another basic question in studies of unresolved galaxies is the slope of 
the IMF. Constraining the IMF requires an access 
to gravity sensitive features and the IR region contains some: the CO 
bands \citep{2004ApJS..151..387I} and the Na{\sc i} doublet at 1.14\,$\mu$m
\citep{2015MNRAS.454L..71S}. Their behaviour suffers from a degeneracy 
between the IMF slope and metallicity, so more complex approach of
solving the stellar populations is needed here.

Finally, the IR absorption features allow us to probe the kinematics of 
different stellar populations -- \citet{2015MNRAS.446.2823R} found that 
the IR velocity dispersion $\sigma$$_{\rm IR}$ is lower than the optical 
$\sigma$$_{\rm opt}$ and explained this result either with presence of warm 
dust that filled in the CO bands, used to measure $\sigma$$_{\rm IR}$, or 
with contribution from a redder, dynamically colder stellar population 
that dominates the IR light. Disentangling between these possibilities 
needs spatially resolved kinematic measurements, which is within the 
reach of the present state of the art IR instrumentation. 

Addressing these questions motivated some efforts to obtain a large 
high-quality data set of ``normal'' and relatively simple stellar 
systems -- both galaxies and star clusters -- started to attract 
attention as relatively straightforward test cases that can be used to 
gain understanding of the behaviour of IR spectral features and to test 
models. 
\citet{2010A&A...510A..19L,2012A&A...543A..75L} and \citet{2011MNRAS.410.2714R} 
reported integrated spectra over entire star clusters, while
\citet{2008ApJ...674..194S}, \citet{2009A&A...497...41C}, \citet{2012MNRAS.425.1057K} and \citet{2015ApJS..217...13M} 
provided spectra of quiescent galaxies.

Here we present new templates of galaxy spectra with a resolving power 
$R\sim$4000, spanning a wavelength range from 0.3\,$\mu$m to 2.4\,$\mu$m. 
It was obtained with the X-Shooter spectrograph at the European Southern Observatory Very Large telescope (ESO VLT), matching 
the most comprehensive recent stellar spectral library \citep[still 
partially a work in progress, see ][]{2014A&A...565A.117C,2016A&A...589A..36G}.
We provide high-quality test data facilitating the understanding of the 
IR spectral diagnostic tools and the testing of the newest generation 
of spectral synthesis models. Our library spans a combined optical- NIR 
range obtained with the same instrument and sampling the same spatial 
region of the galaxies, 
unlike the previous libraries. Thus we minimize any systematics that 
may originate from sampling of the stellar populations gradients inside 
the galaxies. Our sample is dominated by well-characterized objects with 
known and understood stellar populations, spanning a range of ages and 
metallicities. This is the first paper in the series, and it describes
the data. Indices and other results will be reported in further 
publications.

\section{Galaxy sample selection}\label{sec:sample}

The galaxies in the sample have been selected to be bright 
($B_{\rm t}$$\sim$11$-$13\,mag) and distributed along the Hubble sequence 
from ellipticals to late-type spirals. We preferred  early-type 
galaxies, because they generally represent simpler stellar populations
than the spirals, to facilitate an easier testing of stellar population 
models. We gave further priority to objects with  age  and 
metallicity already available in literature.

The basic properties of our sample of galaxies are reported in Table\,\ref{tab:sample}. Some 
literature data for our galaxies that were used for various comparisons 
throughout this paper are listed in Tables,\ref{tab:lit_vh}, 
\ref{tab:lit_sigma}, \ref{tab:lit_indices1} and \ref{tab:lit_indices2}.
If possible, we gave preference to literature sources that have multiple 
objects in common with our sample.

\begin{table*}
\caption{Parameters of the sample galaxies. The columns show the
following. 
(1): galaxy name; 
(2): morphological classification from Lyon Extragalactic 
Database \citep[HyperLEDA; ][]{2014A&A...570A..13M}; 
(3): numerical morphological type from HyperLEDA; 
(4): major and minor isophotal diameter measured at a surface-brightness 
level of $\mu_{\rm B} = 25$ mag arcsec$^{-2}$ from HyperLEDA; 
(5): radial velocity with respect to the cosmic microwave background 
(CMB) reference frame from HyperLEDA; 
(6): total observed blue magnitude $B_{\rm T}$ from HyperLEDA; 
(7): Galactic extinction in $B$-band from HyperLEDA;
(8): Internal extinction due to inclination in $B$-band from HyperLEDA;
(9): distance obtained as $V_{\rm CMB}/H_0$ with $H_0$=75\,km\,s$^{-1}$\,Mpc$^{-1}$; 
(10): absolute total blue magnitude from $B_{\rm T}$ corrected for extinction 
as in HyperLEDA and adopting $D$;}.
\begin{center}
\begin{small}
\begin{tabular}{lcrccccccc}
\hline
\noalign{\smallskip}
\multicolumn{1}{c}{Galaxy} &
\multicolumn{1}{c}{Type} &
\multicolumn{1}{c}{$T$} &
\multicolumn{1}{c}{$D_{25}\,\times\,d_{25}$} &
\multicolumn{1}{c}{$V_{\rm CMB}$} &
\multicolumn{1}{c}{$B_{\rm T}$} &
\multicolumn{1}{c}{$A_{\rm B}^{\rm g}$} &
\multicolumn{1}{c}{$A_{\rm B}^{\rm i}$} &
\multicolumn{1}{c}{$D$} &
\multicolumn{1}{c}{$M_{B_{\rm T}}$}\\ 
\noalign{\smallskip}
\multicolumn{1}{c}{ID} &
\multicolumn{1}{c}{} &
\multicolumn{1}{c}{} &
\multicolumn{1}{c}{[arcmin]} &
\multicolumn{1}{c}{[\kms]} &
\multicolumn{1}{c}{[mag]} &
\multicolumn{1}{c}{[mag]} &
\multicolumn{1}{c}{[mag]} &
\multicolumn{1}{c}{[Mpc]} &
\multicolumn{1}{c}{[mag]}  \\
\noalign{\smallskip}
\multicolumn{1}{c}{(1)} &
\multicolumn{1}{c}{(2)} &
\multicolumn{1}{c}{(3)} &
\multicolumn{1}{c}{(4)} &
\multicolumn{1}{c}{(5)} &
\multicolumn{1}{c}{(6)} &
\multicolumn{1}{c}{(7)} &
\multicolumn{1}{c}{(8)} &
\multicolumn{1}{c}{(9)} &
\multicolumn{1}{c}{(10)}  \\
\noalign{\smallskip}
\hline
\noalign{\smallskip}  
NGC\,0584 &E    & $-$4.7 & 3.80$\times$2.51 & 1558 & 11.33 & 0.18 & 0.00 & 20.7 & $-$20.43    \\
NGC\,0636 &E    & $-$4.9 & 2.69$\times$2.29 & 1570 & 12.35 & 0.11 & 0.00 & 20.9 & $-$19.36    \\
NGC\,0897 &Sa   &    1.1 & 1.99$\times$1.41 & 4572 & 12.81 & 0.09 & 0.14 & 61.0 & $-$21.35   \\
NGC\,1357 &Sab  &    1.9 & 3.38$\times$2.57 & 1861 & 12.40 & 0.19 & 0.15 & 24.8 & $-$19.91   \\
NGC\,1425 &Sb   &    3.2 & 4.78$\times$2.09 & 1404 & 11.32 & 0.06 & 0.53 & 18.7 & $-$20.63    \\
NGC\,1600 &E    & $-$4.6 & 3.09$\times$2.00 & 4649 & 11.94 & 0.19 & 0.00 & 62.0 & $-$22.21    \\
NGC\,1700 &E    & $-$4.7 & 3.09$\times$1.94 & 3862 & 12.03 & 0.19 & 0.00 & 51.4 & $-$21.71    \\
NGC\,2613 &Sb   &    3.1 & 7.85$\times$1.77 & 1940 & 11.11 & 0.38 & 0.85 & 25.8 & $-$22.18   \\
NGC\,3115 &E/S0 & $-$2.9 & 7.07$\times$3.01 &  991 & 10.08 & 0.20 & 0.00 & 13.2 & $-$20.72    \\
NGC\,3377 &E    & $-$4.8 & 3.89$\times$1.86 & 1018 & 11.13 & 0.15 & 0.00 & 13.5 & $-$19.67    \\
NGC\,3379 &E    & $-$4.8 & 4.89$\times$4.26 & 1247 & 10.23 & 0.11 & 0.00 & 16.6 & $-$20.98    \\
NGC\,3423 &Sc   &    6.0 & 3.54$\times$3.01 & 1354 & 11.61 & 0.13 & 0.12 & 18.0 & $-$19.92    \\
NGC\,4415 &S0/a & $-$0.9 & 1.28$\times$1.14 & 1240 & 13.59 & 0.09 & 0.00 & 16.5 & $-$17.59    \\
NGC\,7424 &Sc   &    6.0 & 5.01$\times$2.69 &  692 & 11.05 & 0.05 & 0.47 & 9.22 & $-$19.29    \\
\noalign{\smallskip}
\hline
\noalign{\medskip}
\end{tabular}
\end{small}
\label{tab:sample}
\end{center}
\end{table*}

\section{Observations and data reduction}\label{sec:obs_data}

The data were acquired with the X-Shooter \citep{2006SPIE.6273E..3RG},
a spectrograph at the Unit 2 of ESO Very Large Telescope on Paranal
(Chile). This is a three arms cross-dispersed spectrograph that
provides continuous coverage over $\lambda$$\sim$360-2500\,nm. We used
1.3, 1.5 and 1.2\,arcsec wide slits for the ultraviolet (UBV), visual
(VIS) and near-infrared (NIR) arms, yielding average resolving power
of about 4000, 5400 and 4300, respectively.  The binning in the UBV
and VIS arms was 1$\times$1.

We observed the sample of galaxies along their major axis with the
  aim to derive also the values of the stellar populations properties
  in the center of the galaxy and to evaluate their gradient along the
  major axis. In the final spectra even if the S/N was very high it
  was not enough to measure the gradients of the stellar populations
  properties in the NIR spectral region and we finally co-added the spectra
  along the spatial axis in order to increase the final S/N. This
  results in  spectra mapping a region of about 1.5$\times$1.5 arcsec
  around the center that covers an area ranging values between between  65
  and 430 pc.
 
For all the targets, multiple exposures have been acquired to remove
better the detectors' cosmetics. The  NIR observations were further
split to remove the sky emission and avoid saturation of the brightest
sky lines. For most galaxies we sampled clear sky offsetting away from
the target, except for a few more compact ones for which we observed
in stare mode. A summary of the observations is presented in
Table\,\ref{tab:obs_log}.

\begin{table*}[]
\caption{Observing log. All data were obtained under thin cloud
conditions. Multiple entries for some galaxies indicate that they were 
observed more than once. Here N$_{exp}$ is the number of exposures when 
the target was on the slit, regardless of the exact observing strategy, 
e.g. staring or generic offset.}\label{tab:obs_log}
\begin{center}
\begin{small}
\begin{tabular}{@{ }l@{ }c@{ }c@{ }c@{ }c@{ }c@{ }c@{ }c@{ }c@{ }c@{ }c@{ }}
\hline\hline
\multicolumn{1}{c}{Galaxy} & Observing start  & \multicolumn{3}{c}{S/N} & \multicolumn{3}{c}{Total Exposure times on target [s]}                      & sec\,$z$ & Seeing        & Sky  \\
\multicolumn{1}{c}{ID}     & UT date          & 450  & 650  & 2200                  & UV (N$_{\rm exp}$\,$\times$ & VIS (N$_{\rm exp}$\,$\times$ & NIR (N$_{exp}$\,$\times$ &      & FWHM         & Mode \\
                           & [yyyy-mm-ddThh:mm] & [nm] & [nm] & [nm]                  & \,Integration)          & \,Integration)           & NDIT\,$\times$\,DIT)     &          & [arcsec] &      \\
\hline
NGC\,0584   & 2010-10-16T01:06 &  80 & 100 & 120 &  460\,(2$\times$230) &  276\,(2$\times$138) &  600\,(2$\times$1$\times$300)~&~1.66--1.50 & ~0.6 & offset \\
NGC\,0636   & 2010-11-16T00:36 &  80 & 100 & 120 &  690\,(3$\times$230) &  414\,(3$\times$138) &  900\,(3$\times$1$\times$300)~&~1.21--1.13 & ~1.3 & offset \\
NGC\,0897   & 2010-10-02T03:28 &  60 & 120 &  75 & 1060\,(2$\times$530) &  876\,(2$\times$438) & 1200\,(2$\times$2$\times$300)~&~1.27--1.15 & ~1.9 & offset \\
NGC\,1357   & 2010-11-16T01:25 &  70 & 100 &  90 & 1280\,(2$\times$640) & 1100\,(2$\times$550) & 1440\,(2$\times$3$\times$240)~&~1.44--1.22 & ~1.7 & stare  \\
NGC\,1425 a & 2010-10-06T06:27 & 100 & 120 &  85 & 2120\,(4$\times$530) &  876\,(2$\times$438) & 1200\,(2$\times$2$\times$300)~&~1.03--1.01 & ~1.8 & offset \\
NGC\,1425 b & 2010-11-17T02:07 &  80 & 110 & 100 & 1060\,(2$\times$530) &  876\,(2$\times$438) & 1200\,(2$\times$2$\times$300)~&~1.21--1.11 & ~0.9 & offset \\
NGC\,1600   & 2010-11-25T07:37 &  40 &  80 &  50 & 1060\,(2$\times$530) &  876\,(2$\times$438) & 1200\,(2$\times$2$\times$300)~&~1.11--1.19 & ~1.7 & offset \\
NGC\,1700   & 2010-11-21T03:49 &  70 & 100 & 160 &  690\,(3$\times$230) &  414\,(3$\times$138) &  900\,(3$\times$1$\times$300)~&~1.19--1.12 & ~2.5 & offset \\
NGC\,2613   & 2010-11-25T07:16 &  40 &  90 & 108 & 1280\,(2$\times$640) & 1100\,(2$\times$550) & 1440\,(2$\times$3$\times$240)~&~1.09--1.02 & ~2.3 & offset \\
NGC\,3115   & 2011-02-15T01:41 & 140 & 190 & 145 &  690\,(3$\times$230) &  414\,(3$\times$138) &  900\,(3$\times$1$\times$300)~&~1.63--1.45 & ~1.1 & stare  \\  
NGC\,3377   & 2011-03-01T05:41 & 160 & 200 & 160 & 1060\,(2$\times$530) &  876\,(2$\times$438) & 1200\,(2$\times$2$\times$300)~&~1.31--1.39 & ~2.1 & offset \\
NGC\,3379   & 2011-03-06T04:19 & 100 & 110 & 153 &  460\,(2$\times$230) &  278\,(2$\times$138) &  600\,(2$\times$1$\times$300)~&~1.26--1.25 & ~3.0 & offset \\
NGC\,3423   & 2011-03-22T06:25 &  10 &  20 &  10 & 1280\,(2$\times$640) & 1100\,(2$\times$550) & 1440\,(2$\times$3$\times$240)~&~1.59--2.08 & ~0.8 & offset \\
NGC\,4415 a & 2011-02-18T05:00 &  10 &  20 &  10 & 1060\,(2$\times$530) &  876\,(2$\times$438) & 1200\,(2$\times$2$\times$300)~&~1.45--1.31 & ~1.0 & offset \\
NGC\,4415 b & 2011-02-18T05:43 &  10 &  20 &  12 & 1060\,(2$\times$530) &  876\,(2$\times$438) & 1200\,(2$\times$2$\times$300)~&~1.31--1.23 & ~1.1 & offset \\
NGC\,7424 a & 2010-11-15T00:57 &  30 &  45 &   6 &  640\,(1$\times$640) &  550\,(1$\times$550) & 1440\,(2$\times$3$\times$240)~&~1.06--1.12 & ~0.6 & stare  \\
NGC\,7424 b & 2010-11-15T02:08 &  30 &  45 &   7 & 2560\,(4$\times$640) & 1100\,(4$\times$550) & 1440\,(2$\times$3$\times$240)~&~1.16--1.30 & ~0.9 & stare  \\
\hline
\end{tabular}
\end{small}
\end{center}
\end{table*}

The spectra were reduced using the X-Shooter pipeline ver. 2.6.8.
\citep{2006SPIE.6269E..2KG} in the ESO Reflex workflow environment 
ver. 2.8.1 \citep{2013A&A...559A..96F}.
The basic processing steps are, bias and background subtraction, 
cosmic ray removal \citep{2001PASP..113.1420V}, sky subtraction 
\citep{2003PASP..115..688K}, flat-fielding, order extraction, and 
merging. The telluric corrections was performed with {\it molecfit}, 
a software tool designed to remove atmospheric absorption features 
from astronomical spectra \citep{2015A&A...576A..77S}.

The 1-dimensional spectrum extraction was performed over a region 
of $\pm$2 \arcsec from the peak value of the spectrum. A flux 
correction was applied to take into account the regions outside 
these limits by fitting a Gaussian profile to the cross order 
profile and integrating over the fit to obtain the total flux. The
spectra from the three X-Shooter arms were combined matching them 
to the bluest spectrum, so the UBV arm is anchoring the overall 
flux calibration.

\begin{figure}
\centering
\includegraphics[angle=0.0,width=7.8cm]{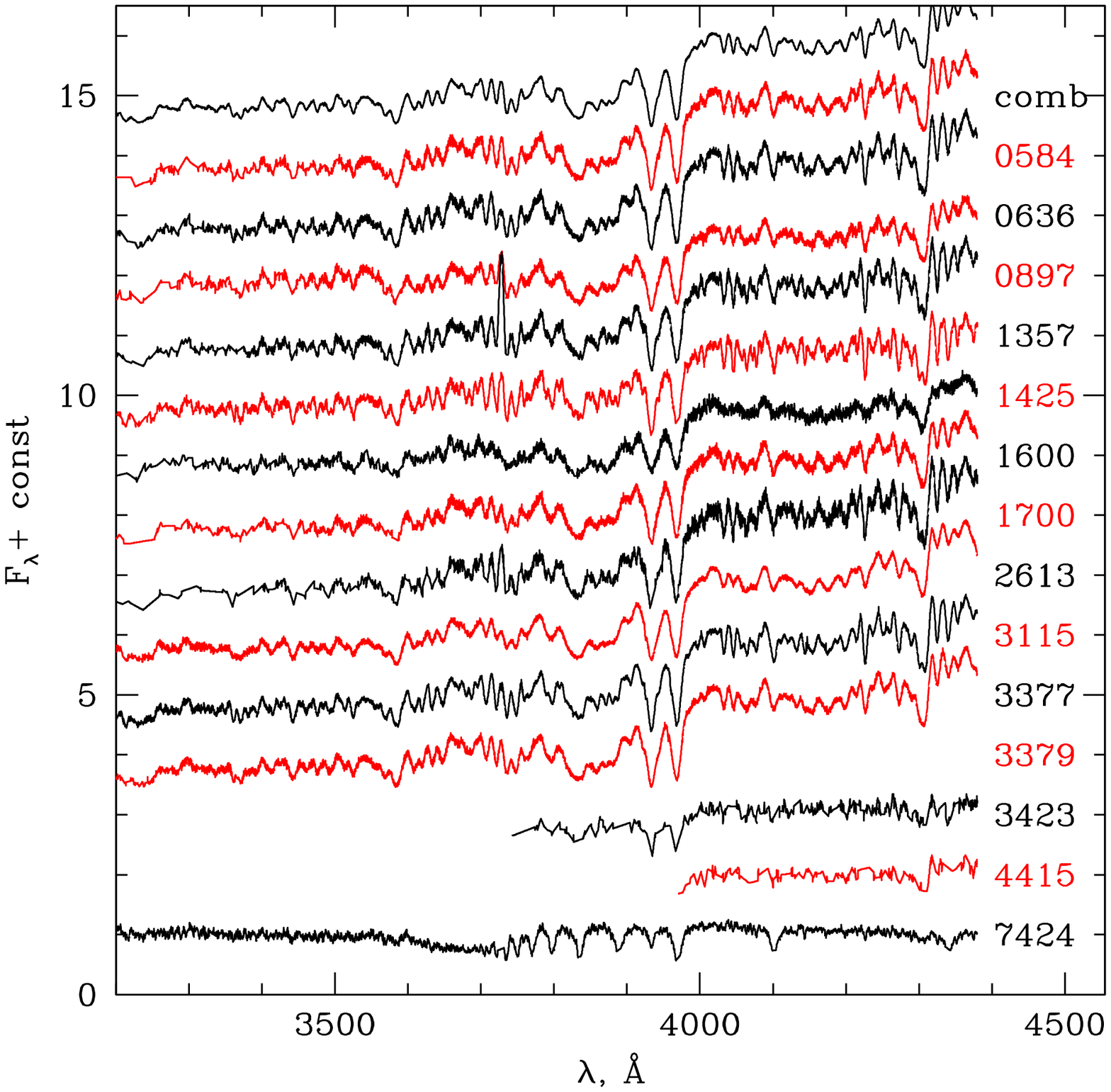}\\
\includegraphics[angle=0.0,width=7.8cm]{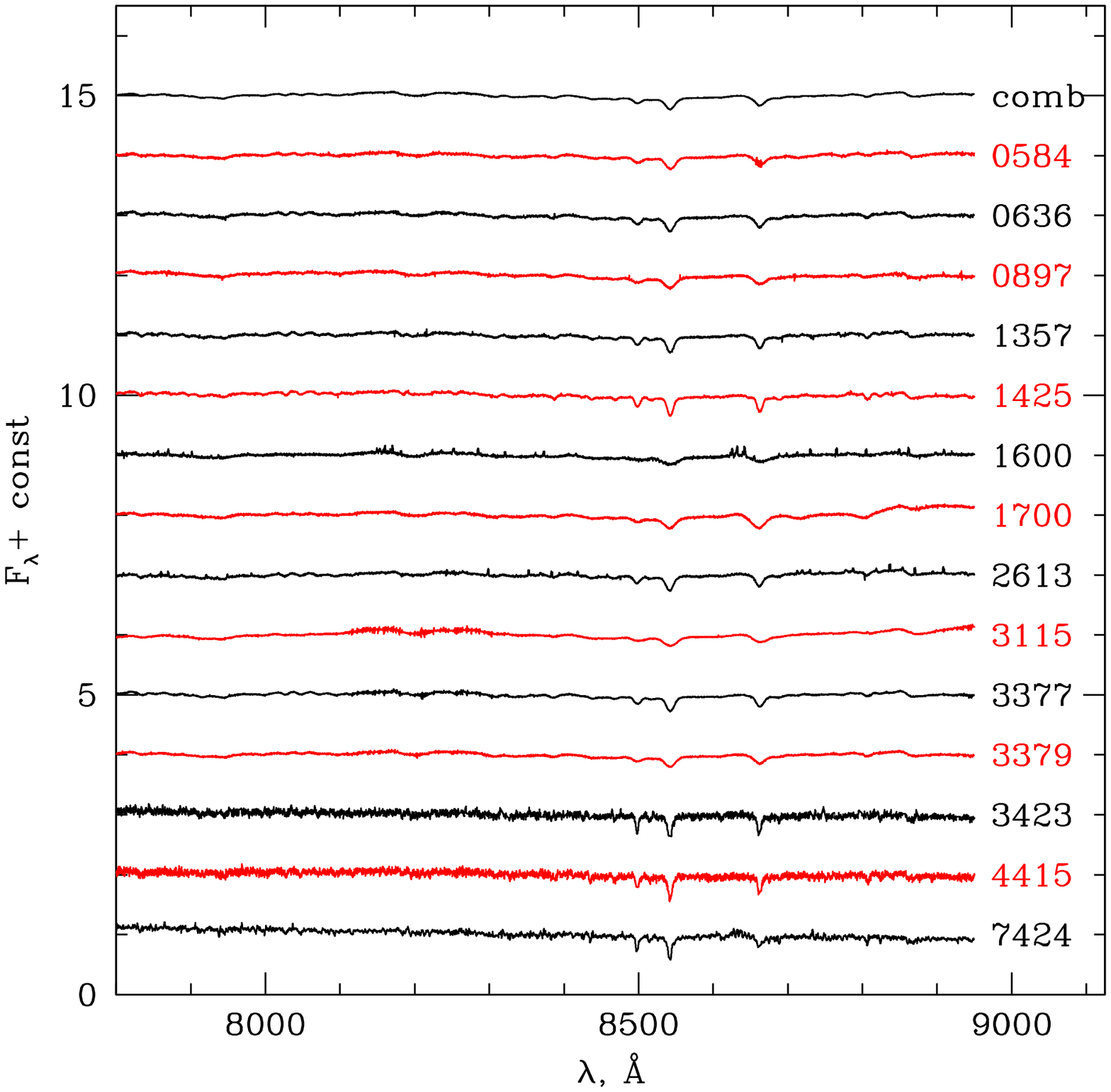}\\
\includegraphics[angle=0.0,width=7.8cm]{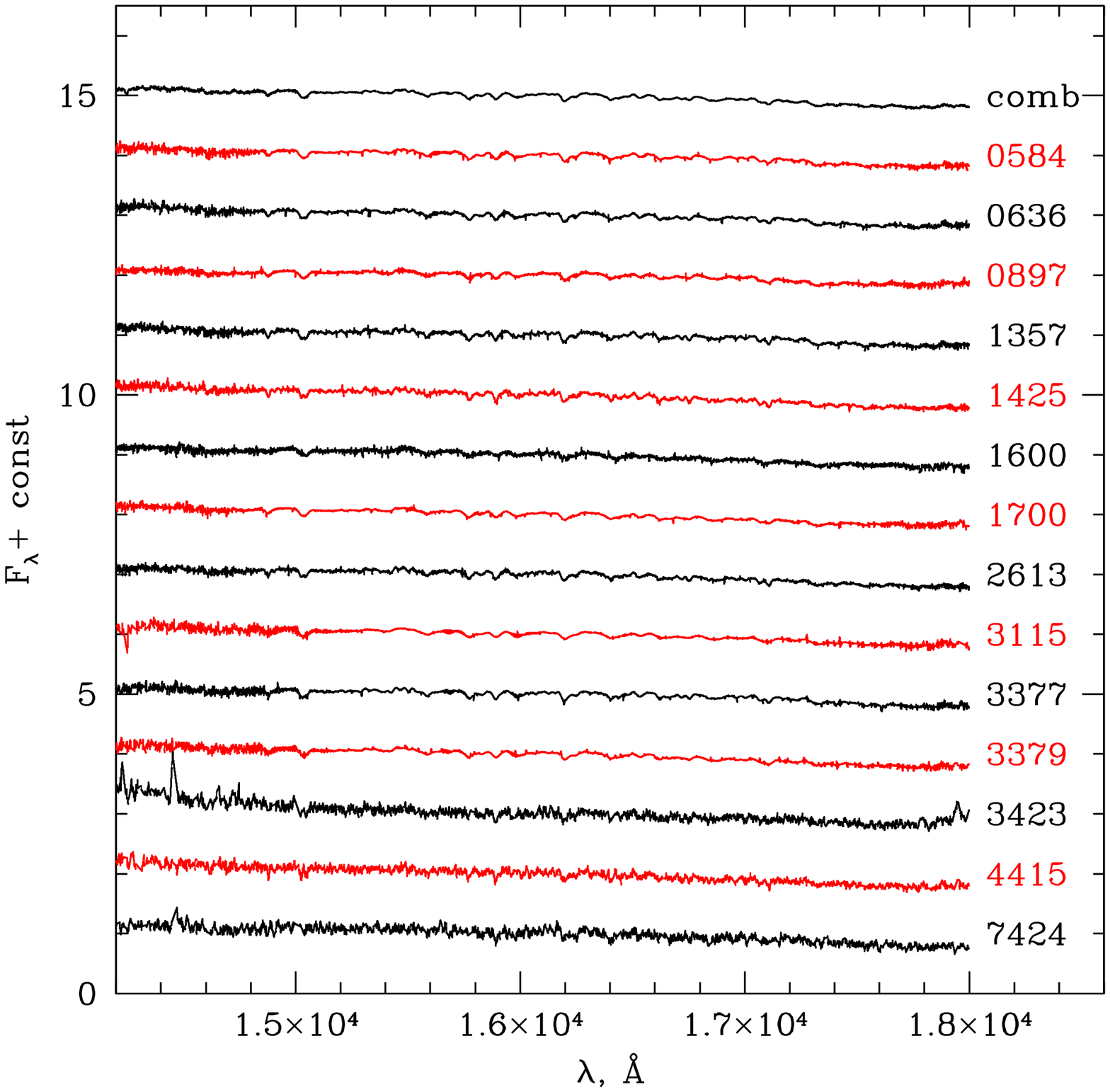}\\
\caption{Spectra of the sample galaxies in different wavelength ranges. Each spectrum is normalized to unity, shifted for displaying purposes and labelled with the NGC number of the galaxy. The spectra on the top of each panel are the average of the sample spectra. (see Sec.\,\ref{sec:obs_data}).}
\label{fig:spectra1}
\end{figure}

The final products are shown in Fig.\ref{fig:spectra1} and 
Fig.\ref{fig:spectra2}. A flux-weighted combined spectrum with 
3$\sigma$ rejection is also plotted.

\section{Data quality validation}\label{sec:validation}

First, to evaluate the internal consistency of the galaxy spectra we 
compared the products generated from different observations of the same 
galaxy (Fig\,\ref{fig:spectra_comp}).  For example, the $r.m.s.$ of the difference over 
14-20\,nm long intervals are $\sim$1.5\,\% and 8-10\,\%, respectively 
for NGC\,1425 and NGC\,4415. The resulting values reflect the different S/N of the spectra, 
while the flat residuals are an  indication a lack of  systematic effects.

\begin{figure}
\centering
\includegraphics[angle=0.0,width=8.8cm]{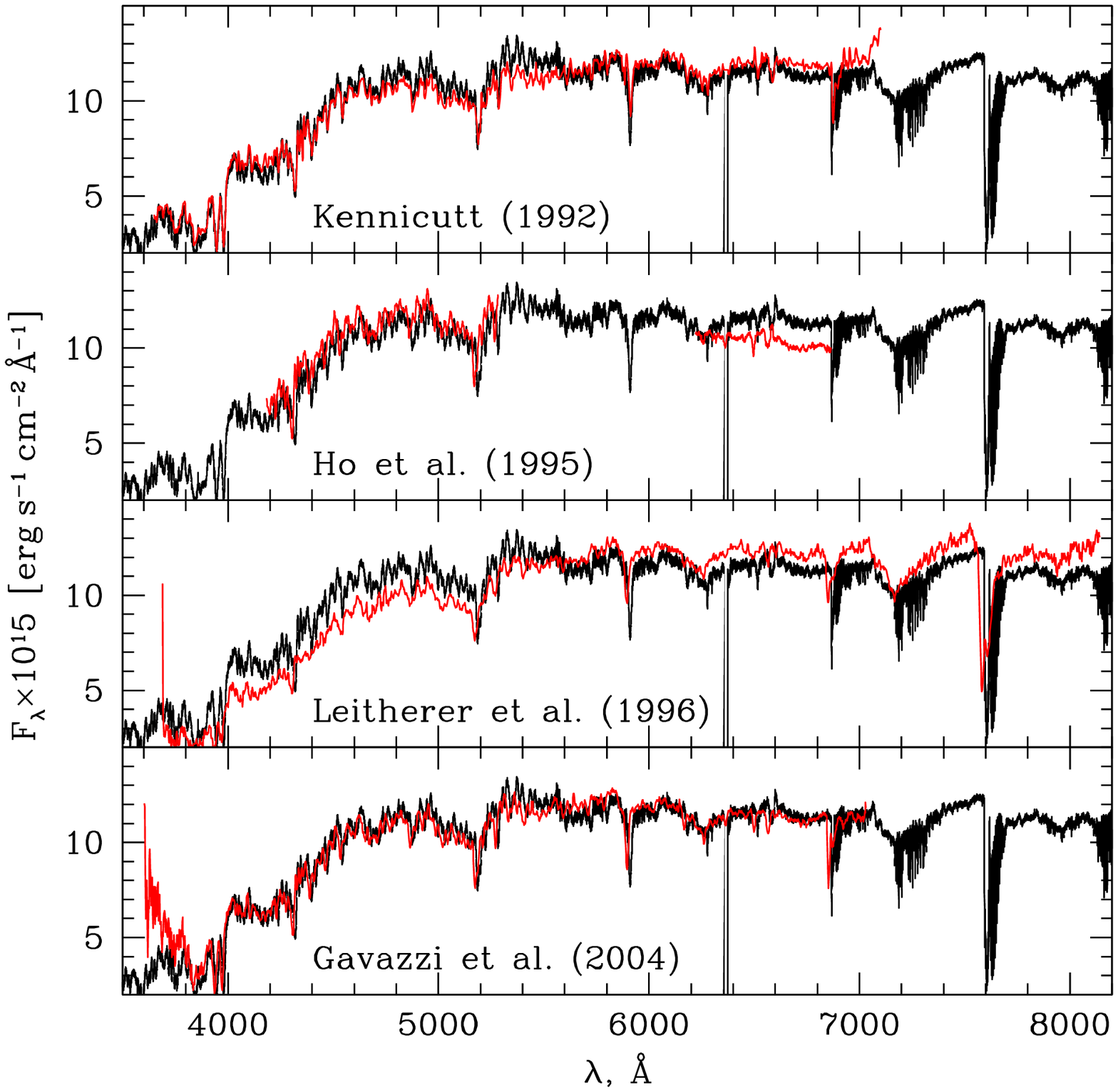}\\
\includegraphics[angle=0.0,width=8.8cm]{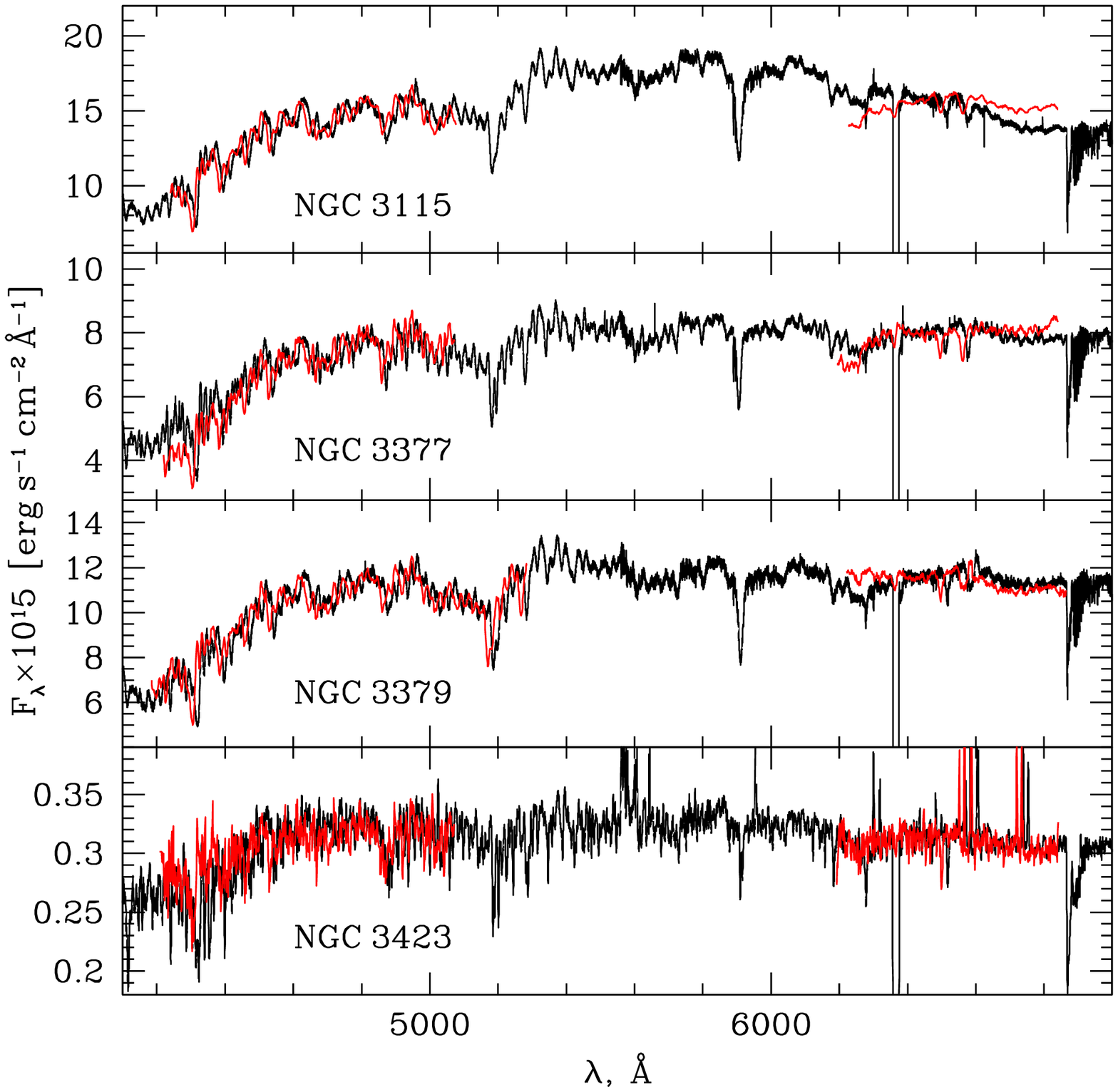}\\
\caption{
Comparison between our optical spectra (black lines) with literature spectra (red lines) scaled to match the flux level of ours. {\em Top panels}: Spectra of NGC 3379 from different sources as given in the labels. {\em Bottom panels}: Spectra of the four galaxies in common with \citet{1995ApJS...98..477H} . Their spectra consist of two portions, which we separately scaled to match our spectrum }

\label{fig:spectra_comp_calibrated}
\end{figure}

Next, a number of optical spectra of the galaxies in our sample are 
available from the literature (Table\,\ref{tab:spectra_literature}). 
They provide an external check. Unfortunately, the largest overlap with 
8 objects in common with the 6dF survey  \citep{2009MNRAS.399..683J} is 
not useful, because their spectra were obtained with a fiber spectrograph 
and not flux calibrated (Fig.\,\ref{fig:spectra_comp_J09}). Comparisons 
with some flux calibrated spectra from other sources are plotted in 
Fig.\,\ref{fig:spectra_comp_calibrated} and show reasonably good matches. 
The most notable deviations are in the overall slopes.

Unfortunately, there is no overlap between our sample and those of the 
more modern NIR spectroscopic surveys of galaxies
\citep[e.g., ][]{2008ApJ...674..194S,2009A&A...497...41C,2012MNRAS.425.1057K,2012ApJS..203...14M,2015ApJS..217...13M}.
NGC\,3379 was observed by \citet{2000ApJ...545..190I}, but only in the 
$K$ atmospheric window, and with low S/N$\sim$50 and an 
resolving power of only $ R \sim$1200. \citet{2009A&A...497...41C} 
published an average low-resolution ($R \sim$1000) NIR spectrum of 
elliptical galaxies. A direct comparison with the NIR spectra of five 
bonafide ellipticals show up to 15\,\% deviations over a single 
atmospheric window.

Following the example of other spectral libraries, we attempted 
to evaluate how well our data processing recovers the shape of the spectra. 
This issue is particularly important for data taken with cross-dispersed 
spectrographs, which implies that the final spectrum is assembled from 
multiple, essentially independently reduced spectral orders. 
\citet{2009ApJS..185..289R} generated synthetic Two Micron All Sky Survey  
\citep[2MASS;][]{2006AJ....131.1163S}
colors for the stars 
in their sample and compare them with the actual 2MASS observations. 
However, this is a straightforward test only for stellar spectra where the 
slit size and the seeing have limited effect. Furthermore, galaxies may 
have radial stellar population gradients that could drive radial color 
gradients, therefore, it is important to match the apertures used for 
observing in different photometric band-passes both between themselves and 
between them and the slits apertures used to extract our spectra. X-Shooter 
uses slits of different sizes for each of its three arms, but we matched 
the fluxes during the combination (see\,Sec.\,\ref{sec:obs_data}), so the 
flux difference originating from the slit apertures cancels out, to zero 
order -- some residual effect related to the galaxy profiles and color 
gradients may remain.

\begin{figure}
\centering
\includegraphics[angle=0.0,width=7.8cm]{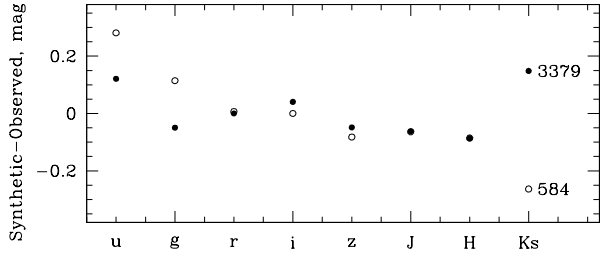}\\
\caption{
 Comparison between the synthetic apparent magnitudes derived from our spectra and observed apparent magnitudes reported  by \citet{2014ApJS..212...18B}
for NGC\,584 (open circles) and NGC\,3379 (solid dots).}
\label{fig:colors}
\end{figure}

A search through the literature identified no homogeneous multi-band and
multi-galaxy flux measurements for our galaxies within apertures with 
sizes as small as a few arcseconds. An exception, and only in terms of 
using an uniform aperture to measure the apparent galaxy magnitudes in 
the same apertures is the work of \citet{2014ApJS..212...18B}. They report
for two of our galaxies (NGC\,584 and NGC\,3379) the  measurements for the Sloan Digital Sky Survey  
\citep[SDSS; ][]{2000AJ....120.1579Y} and the 2MASS 
sets of filters \citep{1996AJ....111.1748F,2003AJ....126.1090C},
although they use wide apertures: 
apertures of $18\times55$ arcsec$^2$ and $180\times120$ arcsec$^2$, respectively.
 To facilitate the comparison 
we formed the difference between the SDSS and 2MASS magnitudes derived 
from our spectra and their apparent magnitudes, and subtracted the median
for each galaxy, effectively removing the aperture effects. The result is
shown in Fig.\,\ref{fig:colors}. The $r.m.s.$ is 0.16\,mag for NGC\,584 and 
0.09\,mag for NGC\,3379, indicating a reasonably good agreement. Removing 
the two largest two outliers -- $u$ and $K_{\rm s}$ -- reduces these values to 
0.08\,mag and 0.05\,mag, respectively, which is comparable to the quality 
of the stellar spectra of \citet{2014ApJS..212...18B}.

The apparent colors for the central $1.3 \times 4$ arcsec$^2$  of the 
galaxies in our sample, derived from the X-Shooter spectra, in some of the 
more commonly used photometric systems are listed Table\,\ref{tab:colors}.

\begin{figure}
\centering
\includegraphics[angle=0.0,width=0.5\textwidth]{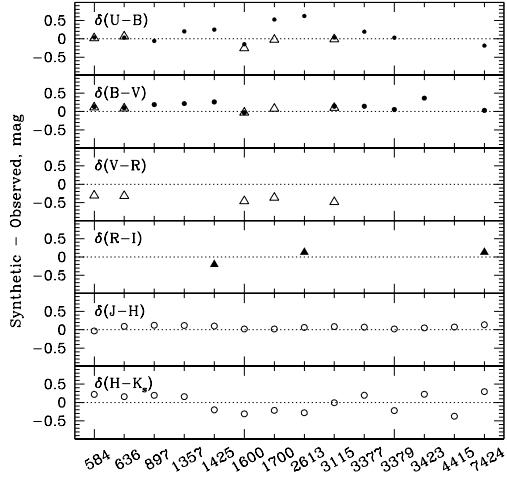}
\caption{A comparison of synthetic broad band colors generated
from our spectra and apparent colors form the literature,
listed in Table \ref{tab:colors}. }
\label{fig:compar}
\end{figure}

\section{Stellar kinematics and line-strength indices}\label{sec:individual}

We measured the line-of-sight velocity distribution of the
stellar component of the sample galaxies from the absorption lines in
the observed wavelength range using the Penalized Pixel Fitting
\citep[{\sc ppxf},][]{capems04} and Gas and Absorption Line Fitting
\citep[{\sc gandalf},][]{sarzetal06} with a code which we adapted
to deal with X-Shooter spectra.  In addition, we simultaneously fitted
the ionized-gas emission lines detected with a $S/N>3$ and we derived
the heliocentric radial velocity $V_{\rm H}$ and velocity dispersion
$\sigma_{\rm v}$, of the stars in the central region of galaxies.  We give the
measured stellar kinematics in Table~\ref{tab:central_values}.

We measured the most commonly used Mg, Fe and \Hb\ line-strength indices of the Lick/IDS 
system \citep{1985ApJS...57..711F, 1994ApJS...94..687W}, the iron index
$\rm{\left<Fe\right> = (Fe5270 + Fe5335)/2}$ of \citet{1990MNRAS.245..217G},
the combined magnesium-iron index 
$[{\rm MgFe}]^{\prime}=\sqrt{{\rm Mg}\,b\,(0.72\times {\rm Fe5270} + 0.28\times{\rm Fe5335})}$
of \citet{2003MNRAS.339..897T} and their errors by following the same 
procedure as in 
\citet{2004MNRAS.354..753M,2015AN....336..208M,2016MNRAS.463.4396M}.
We calculated the central values of \Mgb, \Mgd, \Hb, \Fe, 
\MgFe \/ and the velocity dispersion $\sigma$$_V$, all within 
$0.3\,r_{\rm e}$, where $r_{\rm e}$ is the effective radius of the 
galaxy. They are reported in Table\,\ref{tab:central_values}.
To verify our measurements we compared them with the literature data
both directly and with the model-dependent abundance estimates
$[Z/H]$. The plots show clear and well defined trends.  The most
discrepant values come from NGC\,3423, one of the two galaxies with
the lowest S/N ratio. The $r.m.s.$ of the direct comparison is:
18\,\kms\ for $\sigma$$_V$, 0.14\,\AA\ for \Fe\ and for \Mgd\ and
0.22\,mag for \Mgb, respectively.

We derived the stellar population properties in the centre of the
sample galaxies by comparing the measurements of the line-strength
indices with the model predictions by \citet{2010MNRAS.406..165J} for
the single stellar population as a function of age and metallicity.
We calculated the age and metallicity in the centre of the sample
  galaxies from the central values of line-strength indices given in
  Table \ref{tab:central_values}.  The central values of \Hb\ and
  \MgFe\ are compared with the model predictions by
  \citet{2010MNRAS.406..165J}. In this parameter space the mean age
  and total metallicity appear to be almost insensitive to the
  variations of the $\alpha/$Fe enhancement.
The central mean age and total metallicity of the stellar population
in the centre of the sample galaxies were derived from the values of
line-strength indices given in Table \ref{tab:central_values} by a
linear interpolation between the model points using the iterative
procedure described in \citet{2003A&A...407..423M} and \citet{2008MNRAS.389..341M}.

\begin{figure*}
\centering
\includegraphics[angle=0.0,width=0.9\textwidth]{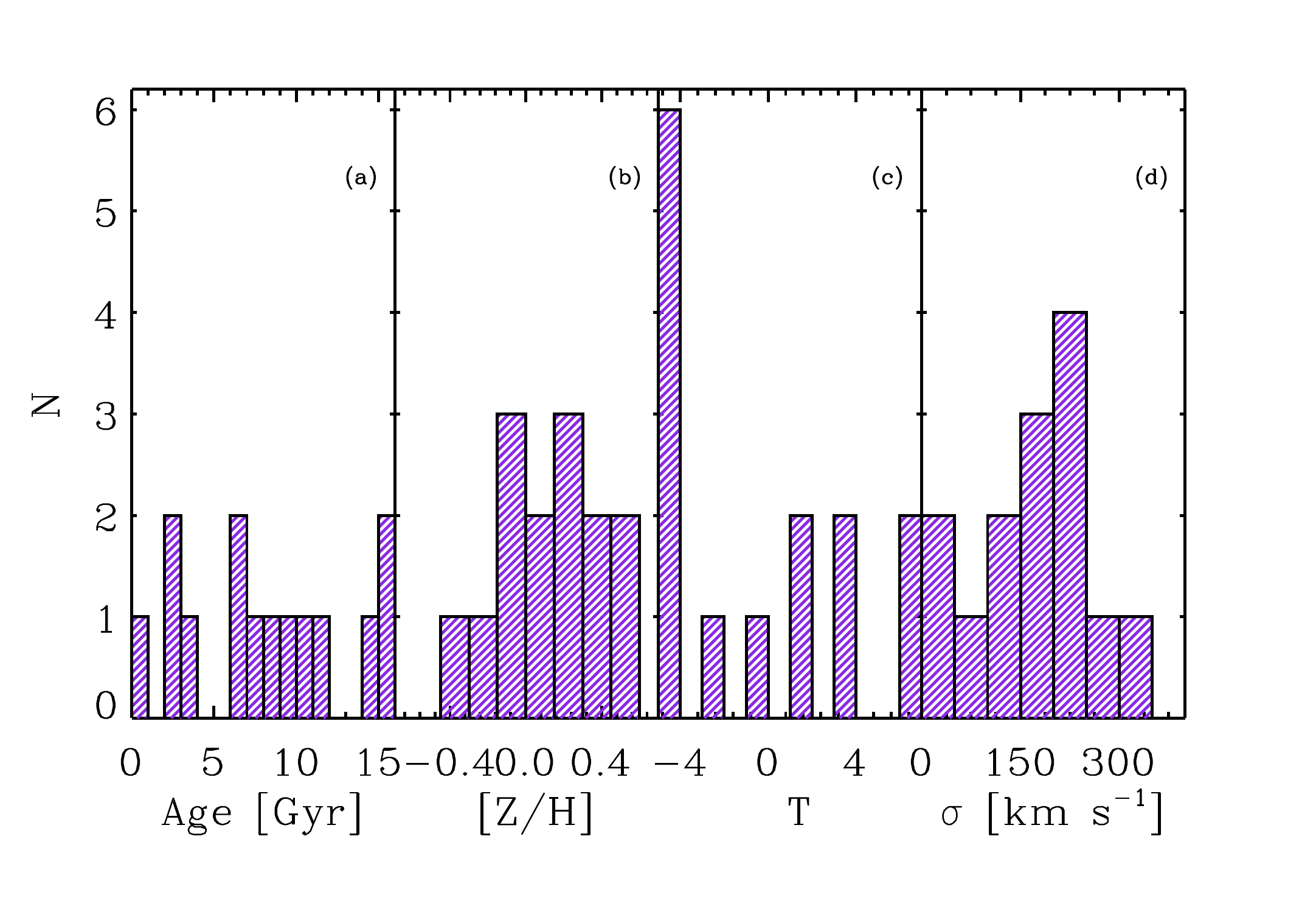}
\caption{Distribution of ages (panel a) and metallicities (panel b)
  morphological type  (panel c) and velocity dispersion  (panel d) for the stellar component in the sample of galaxies.}
\label{fig:istoagemet}
\end{figure*}

We list the central values of age and metallicity in
Table~\ref{tab:sample}.  The histograms of their number
  distribution are plotted in Fig. \ref{fig:istoagemet}. Panel a and
  Panel b of Fig.  \ref{fig:istoagemet} show that the central regions
  of the sample galaxies have a stellar population age and
  metallicity spanning a large range of typical values homogeneously
  distributed.  The total range of values for ages and metallicities
  are respectively $0.8 \leq T \leq 15$ and $\sim -0.5 \leq \ZH \leq
  0.5$.
A detailed study of the history and the evolution of these galaxies based on our results will be presented in a forthcoming paper.

\begin{figure}
\centering
\includegraphics[width=8.6cm]{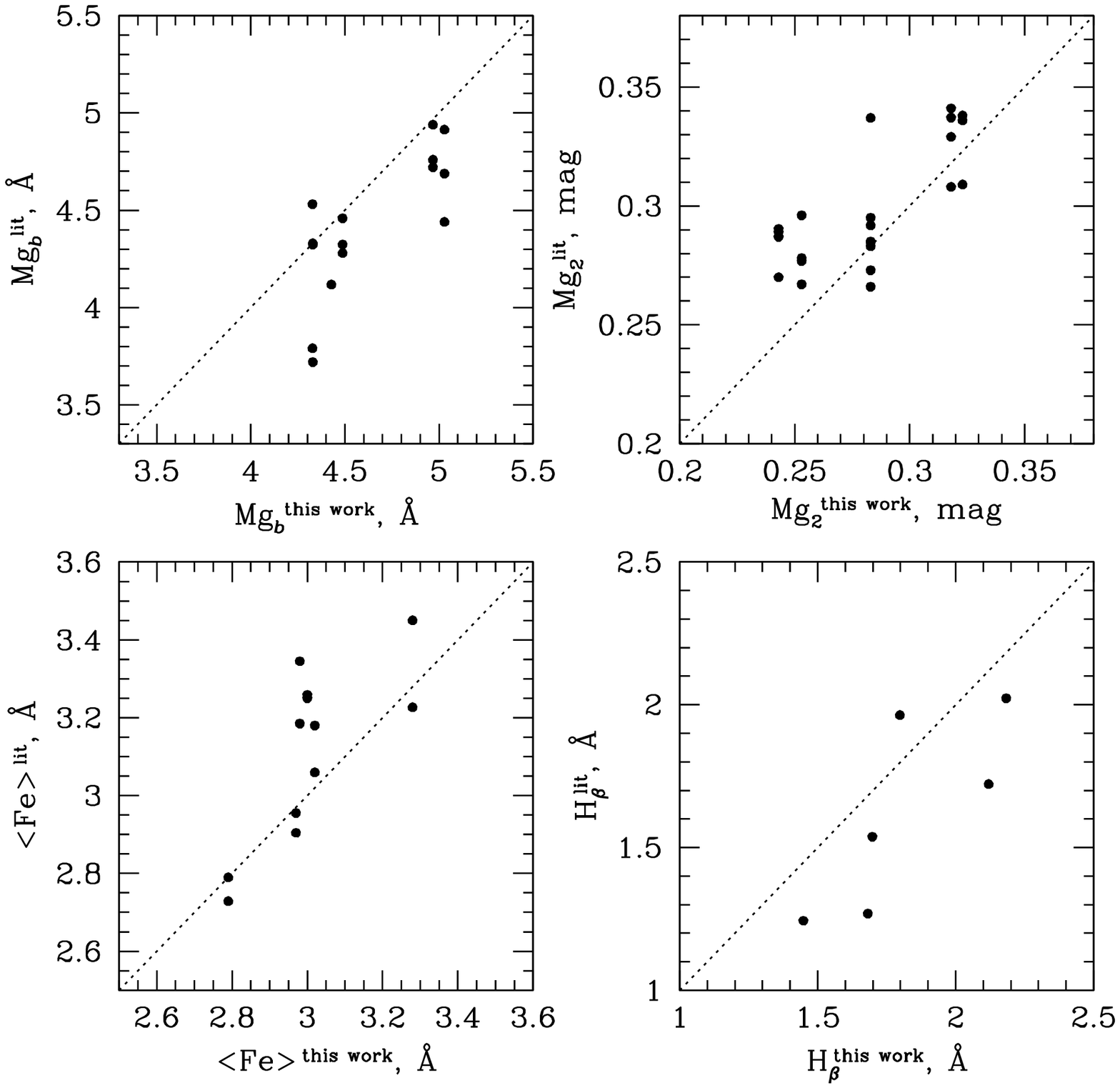} \\
\includegraphics[width=8.6cm]{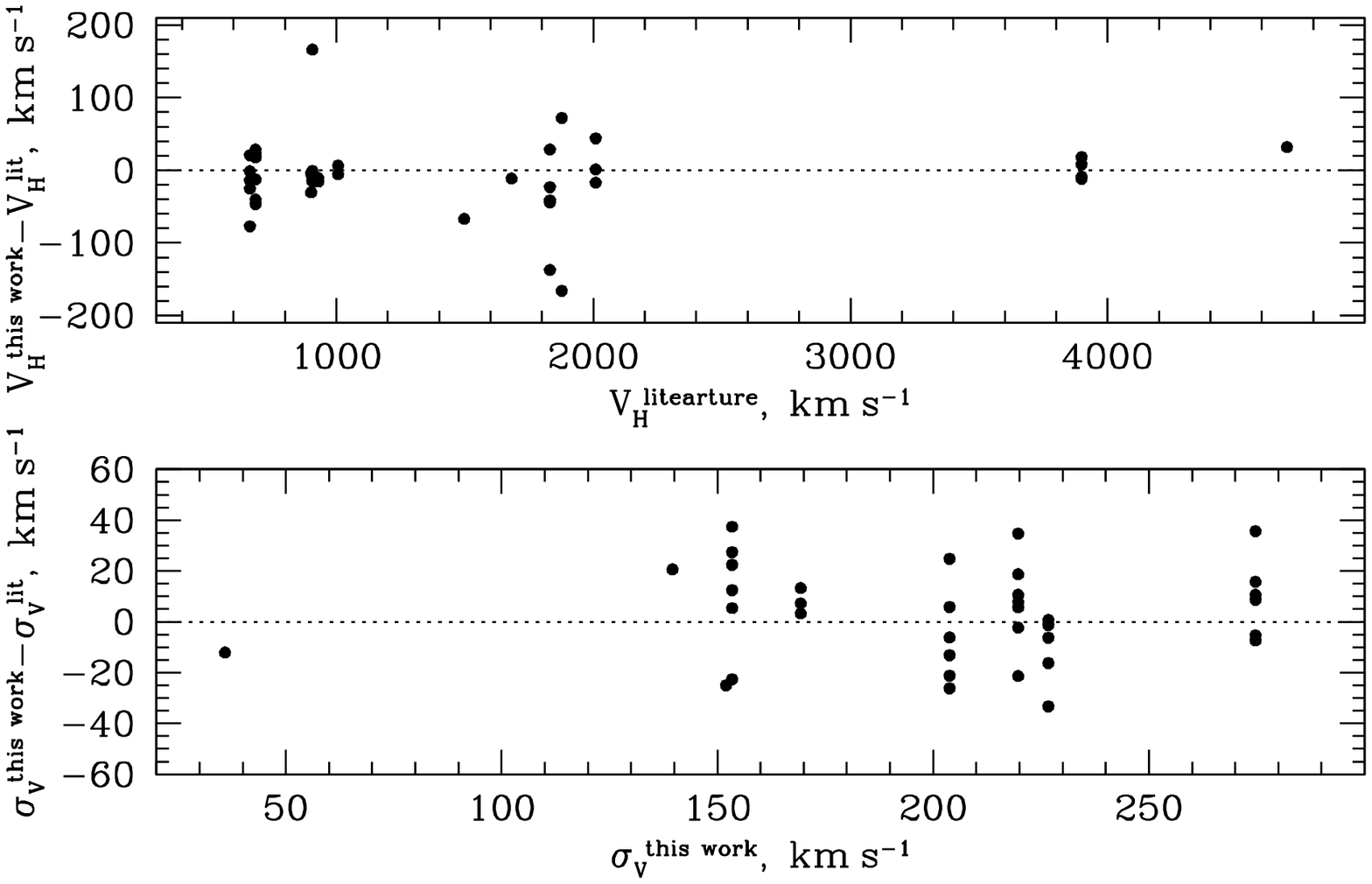}
\caption{{\em Upper panels:} Comparison between our measurements of Lick line-strength indices and those available in literature (Tables\,\ref{tab:central_values} and \ref{tab:lit_vh}
  -- \ref{tab:lit_indices2}). For uniformity we only show literature
  sources that include multiple galaxies from our sample. The plot
  intervals reflect the ranges of values from our measurements.  {\em
    Bottom panels:} 
    Difference between our heliocentric velocities and velocity dispersion in the 
    central region of the sample galaxies and those from literature as a function
     of the literature values. The average difference of 39 measurements after removing three 
     outliers is consistent with zero  $(-$7$\pm$29\,km\,s$^{-1}$).}
\label{fig:indices}
\end{figure}

\begin{table*}
\caption{Central values of the velocity, dispersion, line-strength
  indices, age and metallicity of the sample galaxies measured within an aperture of radius
  $0.3\,r_{\rm e}$.}
\begin{center}
\begin{tiny}
\begin{tabular}{lrrccccccrr}
\hline
\noalign{\smallskip}
\multicolumn{1}{c}{Galaxy} &
\multicolumn{1}{c}{$V_{\rm H}$} &
\multicolumn{1}{c}{$\sigma_V$} &
\multicolumn{1}{c}{\Fe} &
\multicolumn{1}{c}{\MgFe} &
\multicolumn{1}{c}{\Mgd} &
\multicolumn{1}{c}{\Mgb} &
\multicolumn{1}{c}{\Hb} &
\multicolumn{1}{c}{T} &
\multicolumn{1}{c}{\ZH}\\ 
\multicolumn{1}{c}{ID} &
\multicolumn{1}{c}{[\kms]} &
\multicolumn{1}{c}{[\kms]} &
\multicolumn{1}{c}{[\AA]} &
\multicolumn{1}{c}{[\AA]} &
\multicolumn{1}{c}{[mag]} &
\multicolumn{1}{c}{[\AA]} &
\multicolumn{1}{c}{[\AA]} &
\multicolumn{1}{c}{[Gyr]} &
\multicolumn{1}{c}{[dex]} \\
\noalign{\smallskip}
\hline
\noalign{\smallskip}
NGC\,0584 & 1830.6$\pm$1.0 & 203.9$\pm$4.9 & 3.002$\pm$0.143 & 3.696$\pm$0.032 & 0.283$\pm$0.009 & 4.488$\pm$0.145 & 1.698$\pm$0.335 &10.1  $\pm$0.8 &   0.24 $\pm$0.02 \\
NGC\,0636 & 1877.1$\pm$3.8 & 169.3$\pm$3.8 & 3.024$\pm$0.114 & 3.683$\pm$0.021 & 0.283$\pm$0.004 & 4.428$\pm$0.128 & 1.798$\pm$0.243 & 7.7  $\pm$0.8 &   0.32 $\pm$0.05 \\
NGC\,0897 & 4697.9$\pm$1.3 & 225.1$\pm$1.2 & 2.785$\pm$0.125 & 3.695$\pm$0.023 & 0.291$\pm$0.009 & 4.809$\pm$0.183 & 1.289$\pm$0.141 & > 15 $\pm$1.0 &$-$0.06 $\pm$0.02 \\
NGC\,1357 & 2009.0$\pm$3.7 & 139.7$\pm$2.7 & 2.816$\pm$0.138 & 3.529$\pm$0.037 & 0.267$\pm$0.005 & 4.348$\pm$0.163 & 1.907$\pm$0.241 & 6.2  $\pm$0.8 &   0.29 $\pm$0.03 \\
NGC\,1425 & 1497.3$\pm$4.4 & 109.6$\pm$5.4 & 2.642$\pm$0.142 & 3.112$\pm$0.028 & 0.215$\pm$0.006 & 3.598$\pm$0.192 & 1.799$\pm$0.029 & 9.6  $\pm$0.6 &$-$0.01 $\pm$0.02 \\
NGC\,1600 & 4680.8$\pm$5.9 & 335.1$\pm$16.4& 2.853$\pm$0.137 & 3.774$\pm$0.007 & 0.306$\pm$0.003 & 4.912$\pm$0.245 & 1.504$\pm$0.151 &15.0  $\pm$1.8 &   0.12 $\pm$0.04 \\
NGC\,1700 & 3899.2$\pm$2.2 & 226.7$\pm$4.2 & 2.984$\pm$0.148 & 3.636$\pm$0.016 & 0.253$\pm$0.005 & 4.328$\pm$0.161 & 2.183$\pm$0.139 & 2.7  $\pm$0.2 &   0.55 $\pm$0.03 \\
NGC\,2613 & 1681.6$\pm$3.8 & 152.0$\pm$5.9 & 2.835$\pm$0.156 & 3.493$\pm$0.018 & 0.260$\pm$0.002 & 4.231$\pm$0.143 & 1.888$\pm$0.140 & 6.5  $\pm$0.7 &   0.27 $\pm$0.02 \\
NGC\,3115 &  685.6$\pm$2.5 & 274.8$\pm$3.2 & 3.287$\pm$0.146 & 4.102$\pm$0.025 & 0.323$\pm$0.009 & 5.029$\pm$0.200 & 1.682$\pm$0.144 & 8.0  $\pm$1.1 &   0.47 $\pm$0.03 \\
NGC\,3377 &  663.7$\pm$1.6 & 153.4$\pm$4.6 & 2.794$\pm$0.098 & 3.514$\pm$0.018 & 0.243$\pm$0.008 & 4.329$\pm$0.230 & 2.119$\pm$0.129 & 3.5  $\pm$0.1 &   0.42 $\pm$0.01 \\
NGC\,3379 &  907.2$\pm$2.2 & 219.8$\pm$3.2 & 2.975$\pm$0.133 & 3.875$\pm$0.027 & 0.318$\pm$0.002 & 4.968$\pm$0.187 & 1.448$\pm$0.150 & > 15 $\pm$1.7 &   0.15 $\pm$0.04 \\
NGC\,3423 & 1006.3$\pm$2.9 &  38.2$\pm$5.5 & 2.112$\pm$0.125 & 2.153$\pm$0.022 & 0.119$\pm$0.004 & 2.094$\pm$0.127 & 2.758$\pm$0.148 & 2.2  $\pm$0.6 &$-$0.15 $\pm$0.07 \\
NGC\,4415 &  902.6$\pm$2.9 &  35.9$\pm$2.6 & 1.989$\pm$0.243 & 2.405$\pm$0.018 & 0.169$\pm$0.006 & 2.821$\pm$0.169 & 1.865$\pm$0.178 &11.4  $\pm$4.6 &$-$0.39 $\pm$0.15 \\
NGC\,7424 &  929.4$\pm$1.9 &  60.8$\pm$1.4 & 1.489$\pm$0.305 & 1.491$\pm$0.015 & 0.081$\pm$0.007 & 1.383$\pm$0.086 & 4.422$\pm$0.166 & 0.8  $\pm$0.1 &$-$0.22 $\pm$0.05 \\
\hline
\noalign{\bigskip}
\label{tab:central_values}
\end{tabular}
\end{tiny}
\end{center}
\end{table*}

\section{Summary}\label{sec:summary}
 In this paper we presented new high quality UV-VIS-NIR
  spectroscopy of a set of well studied galaxies spanning a large
  range of ages, metallicities and mass, and ranging from ellipticals
  to spirals (Fig. \ref{fig:istoagemet}).This is matching our aim of
  creating an atlas of spectra for a variety of galaxies to be used as
  template to investigate their spectral properties homogeneously from
  the optical to the NIR.

Our X-Shooter dataset of combined  optical-NIR range sampling the
  same spatial region of the galaxies minimize any systematic. The
combination of wide spectral coverage and high S/N in our data is
unique and unprecedented.

Such as a dataset of spectra will be crucial to address important
questions of the modern investigation concerning galaxy formation and
evolution. Some relevant topics among others are the study of the
slope of the IMF, the role of the AGB stars in tracing young stellar
populations, the galaxy continuum and the absorption features in the
IR as complementary tools, with respect to the optical range, in
deriving galaxy stellar populations. These spectra will also be very
extremely useful to be used as templates for stellar population
modeling.

We also discussed a number of tests for the data quality validation to confirm the goodness and consistency of our galaxy spectra.
As consistency check, in Section \ref{sec:validation}, we  compared the spectra of our galaxies with
literature flux calibrated spectra finding a good match between
them.
For two galaxies of our sample, NGC584 and NGC3379, we estimated the quality of
our data processing in recovering the continuum shape of the
spectra. To perform this step we compared synthetic magnitudes
derived from our spectra with observed apparent magnitudes reported by
\citet{2014ApJS..212...18B}. The results indicate a good agreement for
both galaxies.
The flat residuals obtained as results of the ratio of products
generated from different observations of the same galaxy are a strong
indication of a lack of systematic effects in our data.

We measured velocities and velocity
dispersions in the central region of the sample galaxies and we
compared them with literature finding a good consistency within the
scattered literature values reported from several sources.

The classical Lick line-strength indices in the optical range of the X-Shooter
spectra were also measured and their values were
compared with different values from several literature sources. Their
values of  \hb, \Mgb, \Fe\/ agree within the 
uncertainties with the same indices measured in this work. The observed
small offset in the \hb\/ is  due to the different or missing
emission \hb\/ correction in the literature data.

Further in-depth analysis of the UV and the NIR parts of the spectra
will be presented along with their results in a forthcoming paper.

\begin{acknowledgements}
 This paper  made 
extensive use of the SIMBAD Database at CDS (Centre de Donn\'ees 
astronomiques) Strasbourg, the NASA/IPAC Extragalactic Database (NED) 
which is operated by the Jet Propulsion Laboratory, CalTech, under 
contract with NASA, and of the VizieR catalog access tool, CDS, 
Strasbourg, France. PF acknowledges support from the Conseil Scientifique de l'Observatoire de Paris  and the 
Programme National Cosmologie et Galaxies (PNCG).   L.M., A.P., E.M.C. and E.D.B. acknowledge financial support from Padua University through grants
DOR1699945/16, DOR1715817/17, DOR1885254/18, and BIRD164402/16.
\end{acknowledgements}

\bibliographystyle{aa}


\begin{thebibliography}{129}
\expandafter\ifx\csname natexlab\endcsname\relax\def\natexlab#1{#1}\fi

\bibitem[{{Bernardi} {et~al.}(2002){Bernardi}, {Alonso}, {da Costa}, {Willmer},
  {Wegner}, {Pellegrini}, {Rit{\'e}}, \& {Maia}}]{2002AJ....123.2990B}
{Bernardi}, M., {Alonso}, M.~V., {da Costa}, L.~N., {et~al.} 2002, \aj, 123,
  2990

\bibitem[{{Bessell}(1990)}]{1990PASP..102.1181B}
{Bessell}, M.~S. 1990, \pasp, 102, 1181

\bibitem[{{Bonnefoy} {et~al.}(2014){Bonnefoy}, {Chauvin}, {Lagrange}, {Rojo},
  {Allard}, {Pinte}, {Dumas}, \& {Homeier}}]{2014A&A...562A.127B}
{Bonnefoy}, M., {Chauvin}, G., {Lagrange}, A.-M., {et~al.} 2014, \aap, 562,
  A127

\bibitem[{{Bottema}(1989)}]{1989A&A...221..236B}
{Bottema}, R. 1989, \aap, 221, 236

\bibitem[{{Brown} {et~al.}(2014){Brown}, {Moustakas}, {Smith}, {da Cunha},
  {Jarrett}, {Imanishi}, {Armus}, {Brandl}, \& {Peek}}]{2014ApJS..212...18B}
{Brown}, M.~J.~I., {Moustakas}, J., {Smith}, J.-D.~T., {et~al.} 2014, \apjs,
  212, 18

\bibitem[{{Buser} \& {Kurucz}(1992)}]{1992A&A...264..557B}
{Buser}, R. \& {Kurucz}, R.~L. 1992, \aap, 264, 557

\bibitem[{{Cappellari} \& {Emsellem}(2004)}]{capems04}
{Cappellari}, M. \& {Emsellem}, E. 2004, \pasp, 116, 138

\bibitem[{{Castelli} {et~al.}(1997){Castelli}, {Gratton}, \&
  {Kurucz}}]{1997A&A...318..841C}
{Castelli}, F., {Gratton}, R.~G., \& {Kurucz}, R.~L. 1997, \aap, 318, 841

\bibitem[{{Cesetti} {et~al.}(2009){Cesetti}, {Ivanov}, {Morelli}, {Pizzella},
  {Buson}, {Corsini}, {Dalla Bont{\`a}}, {Stiavelli}, \&
  {Bertola}}]{2009A&A...497...41C}
{Cesetti}, M., {Ivanov}, V.~D., {Morelli}, L., {et~al.} 2009, \aap, 497, 41

\bibitem[{{Chen} {et~al.}(2014){Chen}, {Trager}, {Peletier}, {Lan{\c c}on},
  {Vazdekis}, {Prugniel}, {Silva}, \& {Gonneau}}]{2014A&A...565A.117C}
{Chen}, Y.-P., {Trager}, S.~C., {Peletier}, R.~F., {et~al.} 2014, \aap, 565,
  A117

\bibitem[{{Coelho} {et~al.}(2005){Coelho}, {Barbuy}, {Mel{\'e}ndez},
  {Schiavon}, \& {Castilho}}]{2005A&A...443..735C}
{Coelho}, P., {Barbuy}, B., {Mel{\'e}ndez}, J., {Schiavon}, R.~P., \&
  {Castilho}, B.~V. 2005, \aap, 443, 735

\bibitem[{{Cohen} {et~al.}(2003){Cohen}, {Wheaton}, \&
  {Megeath}}]{2003AJ....126.1090C}
{Cohen}, M., {Wheaton}, W.~A., \& {Megeath}, S.~T. 2003, \aj, 126, 1090

\bibitem[{{Colless} {et~al.}(2001){Colless}, {Dalton}, {Maddox}, {Sutherland},
  {Norberg}, {Cole}, {Bland-Hawthorn}, {Bridges}, {Cannon}, {Collins}, {Couch},
  {Cross}, {Deeley}, {De Propris}, {Driver}, {Efstathiou}, {Ellis}, {Frenk},
  {Glazebrook}, {Jackson}, {Lahav}, {Lewis}, {Lumsden}, {Madgwick}, {Peacock},
  {Peterson}, {Price}, {Seaborne}, \& {Taylor}}]{2001MNRAS.328.1039C}
{Colless}, M., {Dalton}, G., {Maddox}, S., {et~al.} 2001, \mnras, 328, 1039

\bibitem[{{Colless} {et~al.}(2003){Colless}, {Peterson}, {Jackson}, {Peacock},
  {Cole}, {Norberg}, {Baldry}, {Baugh}, {Bland-Hawthorn}, {Bridges}, {Cannon},
  {Collins}, {Couch}, {Cross}, {Dalton}, {De Propris}, {Driver}, {Efstathiou},
  {Ellis}, {Frenk}, {Glazebrook}, {Lahav}, {Lewis}, {Lumsden}, {Maddox},
  {Madgwick}, {Sutherland}, \& {Taylor}}]{2003astro.ph..6581C}
{Colless}, M., {Peterson}, B.~A., {Jackson}, C., {et~al.} 2003, ArXiv
  Astrophysics e-prints [\eprint{astro-ph/0306581}]

\bibitem[{{Conroy} \& {van Dokkum}(2012)}]{2012ApJ...747...69C}
{Conroy}, C. \& {van Dokkum}, P. 2012, \apj, 747, 69

\bibitem[{{Cushing} {et~al.}(2005){Cushing}, {Rayner}, \&
  {Vacca}}]{2005ApJ...623.1115C}
{Cushing}, M.~C., {Rayner}, J.~T., \& {Vacca}, W.~D. 2005, \apj, 623, 1115

\bibitem[{{Davies} {et~al.}(2013){Davies}, {Kudritzki}, {Plez}, {Trager},
  {Lan{\c c}on}, {Gazak}, {Bergemann}, {Evans}, \&
  {Chiavassa}}]{2013ApJ...767....3D}
{Davies}, B., {Kudritzki}, R.-P., {Plez}, B., {et~al.} 2013, \apj, 767, 3

\bibitem[{{Davies} {et~al.}(1987){Davies}, {Burstein}, {Dressler}, {Faber},
  {Lynden-Bell}, {Terlevich}, \& {Wegner}}]{1987ApJS...64..581D}
{Davies}, R.~L., {Burstein}, D., {Dressler}, A., {et~al.} 1987, \apjs, 64, 581

\bibitem[{{Doyon} {et~al.}(1994{\natexlab{a}}){Doyon}, {Joseph}, \&
  {Wright}}]{1994ApJ...421..101D}
{Doyon}, R., {Joseph}, R.~D., \& {Wright}, G.~S. 1994{\natexlab{a}}, \apj, 421,
  101

\bibitem[{{Doyon} {et~al.}(1994{\natexlab{b}}){Doyon}, {Wright}, \&
  {Joseph}}]{1994ApJ...421..115D}
{Doyon}, R., {Wright}, G.~S., \& {Joseph}, R.~D. 1994{\natexlab{b}}, \apj, 421,
  115

\bibitem[{{Engelbracht} {et~al.}(1998){Engelbracht}, {Rieke}, {Rieke}, {Kelly},
  \& {Achtermann}}]{1998ApJ...505..639E}
{Engelbracht}, C.~W., {Rieke}, M.~J., {Rieke}, G.~H., {Kelly}, D.~M., \&
  {Achtermann}, J.~M. 1998, \apj, 505, 639

\bibitem[{{Faber} {et~al.}(1985){Faber}, {Friel}, {Burstein}, \&
  {Gaskell}}]{1985ApJS...57..711F}
{Faber}, S.~M., {Friel}, E.~D., {Burstein}, D., \& {Gaskell}, C.~M. 1985,
  \apjs, 57, 711

\bibitem[{{Faber} {et~al.}(1989){Faber}, {Wegner}, {Burstein}, {Davies},
  {Dressler}, {Lynden-Bell}, \& {Terlevich}}]{1989ApJS...69..763F}
{Faber}, S.~M., {Wegner}, G., {Burstein}, D., {et~al.} 1989, \apjs, 69, 763

\bibitem[{{Falco} {et~al.}(1999){Falco}, {Kurtz}, {Geller}, {Huchra}, {Peters},
  {Berlind}, {Mink}, {Tokarz}, \& {Elwell}}]{1999PASP..111..438F}
{Falco}, E.~E., {Kurtz}, M.~J., {Geller}, M.~J., {et~al.} 1999, \pasp, 111, 438

\bibitem[{{Franx} {et~al.}(1989){Franx}, {Illingworth}, \&
  {Heckman}}]{1989ApJ...344..613F}
{Franx}, M., {Illingworth}, G., \& {Heckman}, T. 1989, \apj, 344, 613

\bibitem[{{Freudling} {et~al.}(2013){Freudling}, {Romaniello}, {Bramich},
  {Ballester}, {Forchi}, {Garc{\'{\i}}a-Dabl{\'o}}, {Moehler}, \&
  {Neeser}}]{2013A&A...559A..96F}
{Freudling}, W., {Romaniello}, M., {Bramich}, D.~M., {et~al.} 2013, \aap, 559,
  A96

\bibitem[{{Fukugita} {et~al.}(1996){Fukugita}, {Ichikawa}, {Gunn}, {Doi},
  {Shimasaku}, \& {Schneider}}]{1996AJ....111.1748F}
{Fukugita}, M., {Ichikawa}, T., {Gunn}, J.~E., {et~al.} 1996, \aj, 111, 1748

\bibitem[{{Gavazzi} {et~al.}(2013){Gavazzi}, {Consolandi}, {Dotti}, {Fossati},
  {Savorgnan}, {Gualandi}, \& {Bruni}}]{2013A&A...558A..68G}
{Gavazzi}, G., {Consolandi}, G., {Dotti}, M., {et~al.} 2013, \aap, 558, A68

\bibitem[{{Gavazzi} {et~al.}(2004){Gavazzi}, {Zaccardo}, {Sanvito}, {Boselli},
  \& {Bonfanti}}]{2004A&A...417..499G}
{Gavazzi}, G., {Zaccardo}, A., {Sanvito}, G., {Boselli}, A., \& {Bonfanti}, C.
  2004, \aap, 417, 499

\bibitem[{{Geballe} {et~al.}(1996){Geballe}, {Kulkarni}, {Woodward}, \&
  {Sloan}}]{1996ApJ...467L.101G}
{Geballe}, T.~R., {Kulkarni}, S.~R., {Woodward}, C.~E., \& {Sloan}, G.~C. 1996,
  \apjl, 467, L101

\bibitem[{{Goldoni} {et~al.}(2006){Goldoni}, {Royer}, {Fran{\c c}ois},
  {Horrobin}, {Blanc}, {Vernet}, {Modigliani}, \&
  {Larsen}}]{2006SPIE.6269E..2KG}
{Goldoni}, P., {Royer}, F., {Fran{\c c}ois}, P., {et~al.} 2006, in \procspie,
  Vol. 6269, Society of Photo-Optical Instrumentation Engineers (SPIE)
  Conference Series, 62692K

\bibitem[{{Gonneau} {et~al.}(2016){Gonneau}, {Lan{\c c}on}, {Trager},
  {Aringer}, {Lyubenova}, {Nowotny}, {Peletier}, {Prugniel}, {Chen}, {Dries},
  {Choudhury}, {Falc{\'o}n-Barroso}, {Koleva}, {Meneses-Goytia},
  {S{\'a}nchez-Bl{\'a}zquez}, \& {Vazdekis}}]{2016A&A...589A..36G}
{Gonneau}, A., {Lan{\c c}on}, A., {Trager}, S.~C., {et~al.} 2016, \aap, 589,
  A36

\bibitem[{{Gorgas} {et~al.}(1990){Gorgas}, {Efstathiou}, \& {Aragon
  Salamanca}}]{1990MNRAS.245..217G}
{Gorgas}, J., {Efstathiou}, G., \& {Aragon Salamanca}, A. 1990, \mnras, 245,
  217

\bibitem[{{Gorlova} {et~al.}(2003){Gorlova}, {Meyer}, {Rieke}, \&
  {Liebert}}]{2003ApJ...593.1074G}
{Gorlova}, N.~I., {Meyer}, M.~R., {Rieke}, G.~H., \& {Liebert}, J. 2003, \apj,
  593, 1074

\bibitem[{{Guinouard} {et~al.}(2006){Guinouard}, {Horville}, {Puech}, {Hammer},
  {Amans}, {Chemla}, {Dekker}, \& {Mazzoleni}}]{2006SPIE.6273E..3RG}
{Guinouard}, I., {Horville}, D., {Puech}, M., {et~al.} 2006, in \procspie, Vol.
  6273, Society of Photo-Optical Instrumentation Engineers (SPIE) Conference
  Series, 62733R

\bibitem[{{Hanson} {et~al.}(1997){Hanson}, {Howarth}, \&
  {Conti}}]{1997ApJ...489..698H}
{Hanson}, M.~M., {Howarth}, I.~D., \& {Conti}, P.~S. 1997, \apj, 489, 698

\bibitem[{{Hanson} {et~al.}(2002){Hanson}, {Luhman}, \&
  {Rieke}}]{2002ApJS..138...35H}
{Hanson}, M.~M., {Luhman}, K.~L., \& {Rieke}, G.~H. 2002, \apjs, 138, 35

\bibitem[{{Ho}(2007)}]{2007ApJ...668...94H}
{Ho}, L.~C. 2007, \apj, 668, 94

\bibitem[{{Ho} {et~al.}(1995){Ho}, {Filippenko}, \&
  {Sargent}}]{1995ApJS...98..477H}
{Ho}, L.~C., {Filippenko}, A.~V., \& {Sargent}, W.~L. 1995, \apjs, 98, 477

\bibitem[{{Ho} {et~al.}(2009){Ho}, {Greene}, {Filippenko}, \&
  {Sargent}}]{2009ApJS..183....1H}
{Ho}, L.~C., {Greene}, J.~E., {Filippenko}, A.~V., \& {Sargent}, W.~L.~W. 2009,
  \apjs, 183, 1

\bibitem[{{Huchra} {et~al.}(2012){Huchra}, {Macri}, {Masters}, {Jarrett},
  {Berlind}, {Calkins}, {Crook}, {Cutri}, {Erdo{\v g}du}, {Falco}, {George},
  {Hutcheson}, {Lahav}, {Mader}, {Mink}, {Martimbeau}, {Schneider},
  {Skrutskie}, {Tokarz}, \& {Westover}}]{2012ApJS..199...26H}
{Huchra}, J.~P., {Macri}, L.~M., {Masters}, K.~L., {et~al.} 2012, \apjs, 199,
  26

\bibitem[{{Hyv{\"o}nen} {et~al.}(2009){Hyv{\"o}nen}, {Kotilainen}, {Reunanen},
  \& {Falomo}}]{2009A&A...499..417H}
{Hyv{\"o}nen}, T., {Kotilainen}, J.~K., {Reunanen}, J., \& {Falomo}, R. 2009,
  \aap, 499, 417

\bibitem[{{Imanishi} \& {Alonso-Herrero}(2004)}]{2004ApJ...614..122I}
{Imanishi}, M. \& {Alonso-Herrero}, A. 2004, \apj, 614, 122

\bibitem[{{Ivanov} {et~al.}(2000){Ivanov}, {Rieke}, {Groppi}, {Alonso-Herrero},
  {Rieke}, \& {Engelbracht}}]{2000ApJ...545..190I}
{Ivanov}, V.~D., {Rieke}, G.~H., {Groppi}, C.~E., {et~al.} 2000, \apj, 545, 190

\bibitem[{{Ivanov} {et~al.}(2004){Ivanov}, {Rieke}, {Engelbracht},
  {Alonso-Herrero}, {Rieke}, \& {Luhman}}]{2004ApJS..151..387I}
{Ivanov}, V.~D., {Rieke}, M.~J., {Engelbracht}, C.~W., {et~al.} 2004, \apjs,
  151, 387

\bibitem[{{Johansson} {et~al.}(2010){Johansson}, {Thomas}, \&
  {Maraston}}]{2010MNRAS.406..165J}
{Johansson}, J., {Thomas}, D., \& {Maraston}, C. 2010, \mnras, 406, 165

\bibitem[{{Johnson} \& {M{\'e}ndez}(1970)}]{1970AJ.....75..785J}
{Johnson}, H.~L. \& {M{\'e}ndez}, M.~E. 1970, \aj, 75, 785

\bibitem[{{Jones} {et~al.}(2009){Jones}, {Read}, {Saunders}, {Colless},
  {Jarrett}, {Parker}, {Fairall}, {Mauch}, {Sadler}, {Watson}, {Burton},
  {Campbell}, {Cass}, {Croom}, {Dawe}, {Fiegert}, {Frankcombe}, {Hartley},
  {Huchra}, {James}, {Kirby}, {Lahav}, {Lucey}, {Mamon}, {Moore}, {Peterson},
  {Prior}, {Proust}, {Russell}, {Safouris}, {Wakamatsu}, {Westra}, \&
  {Williams}}]{2009MNRAS.399..683J}
{Jones}, D.~H., {Read}, M.~A., {Saunders}, W., {et~al.} 2009, \mnras, 399, 683

\bibitem[{{Joyce} {et~al.}(1998){Joyce}, {Hinkle}, {Wallace}, {Dulick}, \&
  {Lambert}}]{1998AJ....116.2520J}
{Joyce}, R.~R., {Hinkle}, K.~H., {Wallace}, L., {Dulick}, M., \& {Lambert},
  D.~L. 1998, \aj, 116, 2520

\bibitem[{{Kelson}(2003)}]{2003PASP..115..688K}
{Kelson}, D.~D. 2003, \pasp, 115, 688

\bibitem[{{Kennicutt}(1992)}]{1992ApJS...79..255K}
{Kennicutt}, Jr., R.~C. 1992, \apjs, 79, 255

\bibitem[{{Kleinmann} \& {Hall}(1986)}]{1986ApJS...62..501K}
{Kleinmann}, S.~G. \& {Hall}, D.~N.~B. 1986, \apjs, 62, 501

\bibitem[{{Kotilainen} {et~al.}(2012){Kotilainen}, {Hyv{\"o}nen}, {Reunanen},
  \& {Ivanov}}]{2012MNRAS.425.1057K}
{Kotilainen}, J.~K., {Hyv{\"o}nen}, T., {Reunanen}, J., \& {Ivanov}, V.~D.
  2012, \mnras, 425, 1057

\bibitem[{{Kotilainen} {et~al.}(2001){Kotilainen}, {Reunanen}, {Laine}, \&
  {Ryder}}]{2001A&A...366..439K}
{Kotilainen}, J.~K., {Reunanen}, J., {Laine}, S., \& {Ryder}, S.~D. 2001, \aap,
  366, 439

\bibitem[{{Lambert} {et~al.}(1986){Lambert}, {Gustafsson}, {Eriksson}, \&
  {Hinkle}}]{1986ApJS...62..373L}
{Lambert}, D.~L., {Gustafsson}, B., {Eriksson}, K., \& {Hinkle}, K.~H. 1986,
  \apjs, 62, 373

\bibitem[{{Lan{\c c}on} \& {Mouhcine}(2000)}]{2000ASPC..211...34L}
{Lan{\c c}on}, A. \& {Mouhcine}, M. 2000, in Astronomical Society of the
  Pacific Conference Series, Vol. 211, Massive Stellar Clusters, ed. A.~{Lan{\c
  c}on} \& C.~M. {Boily}, 34

\bibitem[{{Lan{\c c}on} \& {Wood}(2000)}]{2000A&AS..146..217L}
{Lan{\c c}on}, A. \& {Wood}, P.~R. 2000, \aaps, 146, 217

\bibitem[{{Leitherer} {et~al.}(1996){Leitherer}, {Alloin},
  {Fritze-v.~Alvensleben}, {Gallagher}, {Huchra}, {Matteucci}, {O'Connell},
  {Beckman}, {Bertelli}, {Bica}, {Boisson}, {Bonatto}, {Bothun}, {Bressan},
  {Brodie}, {Bruzual}, {Burstein}, {Buser}, {Caldwell}, {Casuso},
  {Cervi{\~n}o}, {Charlot}, {Chavez}, {Chiosi}, {Christian}, {Cuisinier},
  {Dallier}, {de Koter}, {Delisle}, {Diaz}, {Dopita}, {Dorman}, {Fagotto},
  {Fanelli}, {Fioc}, {Garcia-Vargas}, {Girardi}, {Goldader}, {Hardy},
  {Heckman}, {Iglesias}, {Jablonka}, {Joly}, {Jones}, {Kurth}, {Lancon},
  {Lejeune}, {Loxen}, {Maeder}, {Malagnini}, {Marigo}, {Mas-Hesse}, {Meynet},
  {Moller}, {Molla}, {Morossi}, {Nasi}, {Nichols}, {Odegaard}, {Parker},
  {Pastoriza}, {Peletier}, {Robert}, {Rocca-Volmerange}, {Schaerer}, {Schmidt},
  {Schmitt}, {Schommer}, {Schmutz}, {Roos}, {Silva}, {Stasi{\'n}ska},
  {Sutherland}, {Tantalo}, {Traat}, {Vallenari}, {Vazdekis}, {Walborn},
  {Worthey}, \& {Wu}}]{1996PASP..108..996L}
{Leitherer}, C., {Alloin}, D., {Fritze-v.~Alvensleben}, U., {et~al.} 1996,
  \pasp, 108, 996

\bibitem[{{Lejeune} {et~al.}(1997){Lejeune}, {Cuisinier}, \&
  {Buser}}]{1997A&AS..125..229L}
{Lejeune}, T., {Cuisinier}, F., \& {Buser}, R. 1997, \aaps, 125
  [\eprint{astro-ph/9701019}]

\bibitem[{{Lejeune} {et~al.}(1998){Lejeune}, {Cuisinier}, \&
  {Buser}}]{1998A&AS..130...65L}
{Lejeune}, T., {Cuisinier}, F., \& {Buser}, R. 1998, \aaps, 130, 65

\bibitem[{{Lyubenova} {et~al.}(2012){Lyubenova}, {Kuntschner}, {Rejkuba},
  {Silva}, {Kissler-Patig}, \& {Tacconi-Garman}}]{2012A&A...543A..75L}
{Lyubenova}, M., {Kuntschner}, H., {Rejkuba}, M., {et~al.} 2012, \aap, 543, A75

\bibitem[{{Lyubenova} {et~al.}(2010){Lyubenova}, {Kuntschner}, {Rejkuba},
  {Silva}, {Kissler-Patig}, {Tacconi-Garman}, \&
  {Larsen}}]{2010A&A...510A..19L}
{Lyubenova}, M., {Kuntschner}, H., {Rejkuba}, M., {et~al.} 2010, \aap, 510, A19

\bibitem[{{Makarov} {et~al.}(2014){Makarov}, {Prugniel}, {Terekhova},
  {Courtois}, \& {Vauglin}}]{2014A&A...570A..13M}
{Makarov}, D., {Prugniel}, P., {Terekhova}, N., {Courtois}, H., \& {Vauglin},
  I. 2014, \aap, 570, A13

\bibitem[{{Maraston}(2005)}]{2005MNRAS.362..799M}
{Maraston}, C. 2005, \mnras, 362, 799

\bibitem[{{Maraston}(2011)}]{2011ASPC..445..391M}
{Maraston}, C. 2011, in Astronomical Society of the Pacific Conference Series,
  Vol. 445, Why Galaxies Care about AGB Stars II: Shining Examples and Common
  Inhabitants, ed. F.~{Kerschbaum}, T.~{Lebzelter}, \& R.~F. {Wing}, 391

\bibitem[{{Maraston} {et~al.}(2006){Maraston}, {Daddi}, {Renzini}, {Cimatti},
  {Dickinson}, {Papovich}, {Pasquali}, \& {Pirzkal}}]{2006ApJ...652...85M}
{Maraston}, C., {Daddi}, E., {Renzini}, A., {et~al.} 2006, \apj, 652, 85

\bibitem[{{Maraston} \& {Str{\"o}mb{\"a}ck}(2011)}]{2011MNRAS.418.2785M}
{Maraston}, C. \& {Str{\"o}mb{\"a}ck}, G. 2011, \mnras, 418, 2785

\bibitem[{{M{\'a}rmol-Queralt{\'o}} {et~al.}(2009){M{\'a}rmol-Queralt{\'o}},
  {Cardiel}, {S{\'a}nchez-Bl{\'a}zquez}, {Trager}, {Peletier}, {Kuntschner},
  {Silva}, {Cenarro}, {Vazdekis}, \& {Gorgas}}]{2009ApJ...705L.199M}
{M{\'a}rmol-Queralt{\'o}}, E., {Cardiel}, N., {S{\'a}nchez-Bl{\'a}zquez}, P.,
  {et~al.} 2009, \apjl, 705, L199

\bibitem[{{Mason} {et~al.}(2015){Mason}, {Rodr{\'{\i}}guez-Ardila}, {Martins},
  {Riffel}, {Gonz{\'a}lez Mart{\'{\i}}n}, {Ramos Almeida}, {Ruschel Dutra},
  {Ho}, {Thanjavur}, {Flohic}, {Alonso-Herrero}, {Lira}, {McDermid}, {Riffel},
  {Schiavon}, {Winge}, {Hoenig}, \& {Perlman}}]{2015ApJS..217...13M}
{Mason}, R.~E., {Rodr{\'{\i}}guez-Ardila}, A., {Martins}, L., {et~al.} 2015,
  \apjs, 217, 13

\bibitem[{{McElroy}(1995)}]{1995ApJS..100..105M}
{McElroy}, D.~B. 1995, \apjs, 100, 105

\bibitem[{{Mehlert} {et~al.}(2003){Mehlert}, {Thomas}, {Saglia}, {Bender}, \&
  {Wegner}}]{2003A&A...407..423M}
{Mehlert}, D., {Thomas}, D., {Saglia}, R.~P., {Bender}, R., \& {Wegner}, G.
  2003, \aap, 407, 423

\bibitem[{{Melbourne} {et~al.}(2012){Melbourne}, {Williams}, {Dalcanton},
  {Rosenfield}, {Girardi}, {Marigo}, {Weisz}, {Dolphin}, {Boyer}, {Olsen},
  {Skillman}, \& {Seth}}]{2012ApJ...748...47M}
{Melbourne}, J., {Williams}, B.~F., {Dalcanton}, J.~J., {et~al.} 2012, \apj,
  748, 47

\bibitem[{{Meneses-Goytia} {et~al.}(2015{\natexlab{a}}){Meneses-Goytia},
  {Peletier}, {Trager}, {Falc{\'o}n-Barroso}, {Koleva}, \&
  {Vazdekis}}]{2015A&A...582A..96M}
{Meneses-Goytia}, S., {Peletier}, R.~F., {Trager}, S.~C., {et~al.}
  2015{\natexlab{a}}, \aap, 582, A96

\bibitem[{{Meneses-Goytia} {et~al.}(2015{\natexlab{b}}){Meneses-Goytia},
  {Peletier}, {Trager}, \& {Vazdekis}}]{2015A&A...582A..97M}
{Meneses-Goytia}, S., {Peletier}, R.~F., {Trager}, S.~C., \& {Vazdekis}, A.
  2015{\natexlab{b}}, \aap, 582, A97

\bibitem[{{Messineo} {et~al.}(2009){Messineo}, {Davies}, {Ivanov}, {Figer},
  {Schuller}, {Habing}, {Menten}, \& {Petr-Gotzens}}]{2009ApJ...697..701M}
{Messineo}, M., {Davies}, B., {Ivanov}, V.~D., {et~al.} 2009, \apj, 697, 701

\bibitem[{{Meyer} {et~al.}(1998){Meyer}, {Edwards}, {Hinkle}, \&
  {Strom}}]{1998ApJ...508..397M}
{Meyer}, M.~R., {Edwards}, S., {Hinkle}, K.~H., \& {Strom}, S.~E. 1998, \apj,
  508, 397

\bibitem[{{Morelli} {et~al.}(2004){Morelli}, {Halliday}, {Corsini}, {Pizzella},
  {Thomas}, {Saglia}, {Davies}, {Bender}, {Birkinshaw}, \&
  {Bertola}}]{2004MNRAS.354..753M}
{Morelli}, L., {Halliday}, C., {Corsini}, E.~M., {et~al.} 2004, \mnras, 354,
  753

\bibitem[{{Morelli} {et~al.}(2016){Morelli}, {Parmiggiani}, {Corsini},
  {Costantin}, {Dalla Bont{\`a}}, {M{\'e}ndez-Abreu}, \&
  {Pizzella}}]{2016MNRAS.463.4396M}
{Morelli}, L., {Parmiggiani}, M., {Corsini}, E.~M., {et~al.} 2016, \mnras, 463,
  4396

\bibitem[{{Morelli} {et~al.}(2015){Morelli}, {Pizzella}, {Corsini}, {Dalla
  Bont{\`a}}, {Coccato}, {M{\'e}ndez-Abreu}, \&
  {Parmiggiani}}]{2015AN....336..208M}
{Morelli}, L., {Pizzella}, A., {Corsini}, E.~M., {et~al.} 2015, Astronomische
  Nachrichten, 336, 208

\bibitem[{{Morelli} {et~al.}(2008){Morelli}, {Pompei}, {Pizzella},
  {M{\'e}ndez-Abreu}, {Corsini}, {Coccato}, {Saglia}, {Sarzi}, \&
  {Bertola}}]{2008MNRAS.389..341M}
{Morelli}, L., {Pompei}, E., {Pizzella}, A., {et~al.} 2008, \mnras, 389, 341

\bibitem[{{Mould} {et~al.}(2012){Mould}, {Reynolds}, {Readhead}, {Floyd},
  {Jannuzi}, {Cotter}, {Ferrarese}, {Matthews}, {Atlee}, \&
  {Brown}}]{2012ApJS..203...14M}
{Mould}, J., {Reynolds}, T., {Readhead}, T., {et~al.} 2012, \apjs, 203, 14

\bibitem[{{Oppenheimer} {et~al.}(1995){Oppenheimer}, {Kulkarni}, {Matthews}, \&
  {Nakajima}}]{1995Sci...270.1478O}
{Oppenheimer}, B.~R., {Kulkarni}, S.~R., {Matthews}, K., \& {Nakajima}, T.
  1995, Science, 270, 1478

\bibitem[{{Origlia} {et~al.}(1993){Origlia}, {Moorwood}, \&
  {Oliva}}]{1993A&A...280..536O}
{Origlia}, L., {Moorwood}, A.~F.~M., \& {Oliva}, E. 1993, \aap, 280, 536

\bibitem[{{Ramirez} {et~al.}(1997){Ramirez}, {Depoy}, {Frogel}, {Sellgren}, \&
  {Blum}}]{1997AJ....113.1411R}
{Ramirez}, S.~V., {Depoy}, D.~L., {Frogel}, J.~A., {Sellgren}, K., \& {Blum},
  R.~D. 1997, \aj, 113, 1411

\bibitem[{{Rayner} {et~al.}(2009){Rayner}, {Cushing}, \&
  {Vacca}}]{2009ApJS..185..289R}
{Rayner}, J.~T., {Cushing}, M.~C., \& {Vacca}, W.~D. 2009, \apjs, 185, 289

\bibitem[{{Reunanen} {et~al.}(2002){Reunanen}, {Kotilainen}, \&
  {Prieto}}]{2002MNRAS.331..154R}
{Reunanen}, J., {Kotilainen}, J.~K., \& {Prieto}, M.~A. 2002, \mnras, 331, 154

\bibitem[{{Reunanen} {et~al.}(2003){Reunanen}, {Kotilainen}, \&
  {Prieto}}]{2003MNRAS.343..192R}
{Reunanen}, J., {Kotilainen}, J.~K., \& {Prieto}, M.~A. 2003, \mnras, 343, 192

\bibitem[{{Rieke} {et~al.}(1980){Rieke}, {Lebofsky}, {Thompson}, {Low}, \&
  {Tokunaga}}]{1980ApJ...238...24R}
{Rieke}, G.~H., {Lebofsky}, M.~J., {Thompson}, R.~I., {Low}, F.~J., \&
  {Tokunaga}, A.~T. 1980, \apj, 238, 24

\bibitem[{{Rieke} {et~al.}(1988){Rieke}, {Lebofsky}, \&
  {Walker}}]{1988ApJ...325..679R}
{Rieke}, G.~H., {Lebofsky}, M.~J., \& {Walker}, C.~E. 1988, \apj, 325, 679

\bibitem[{{Riffel} {et~al.}(2011{\natexlab{a}}){Riffel}, {Bonatto}, {Cid
  Fernandes}, {Pastoriza}, \& {Balbinot}}]{2011MNRAS.411.1897R}
{Riffel}, R., {Bonatto}, C., {Cid Fernandes}, R., {Pastoriza}, M.~G., \&
  {Balbinot}, E. 2011{\natexlab{a}}, \mnras, 411, 1897

\bibitem[{{Riffel} {et~al.}(2015{\natexlab{a}}){Riffel}, {Mason}, {Martins},
  {Rodr{\'{\i}}guez-Ardila}, {Ho}, {Riffel}, {Lira}, {Gonzalez Martin},
  {Ruschel-Dutra}, {Alonso-Herrero}, {Flohic}, {McDermid}, {Ramos Almeida},
  {Thanjavur}, \& {Winge}}]{2015MNRAS.450.3069R}
{Riffel}, R., {Mason}, R.~E., {Martins}, L.~P., {et~al.} 2015{\natexlab{a}},
  \mnras, 450, 3069

\bibitem[{{Riffel} {et~al.}(2006){Riffel}, {Rodr{\'{\i}}guez-Ardila}, \&
  {Pastoriza}}]{2006A&A...457...61R}
{Riffel}, R., {Rodr{\'{\i}}guez-Ardila}, A., \& {Pastoriza}, M.~G. 2006, \aap,
  457, 61

\bibitem[{{Riffel} {et~al.}(2011{\natexlab{b}}){Riffel}, {Ruschel-Dutra},
  {Pastoriza}, {Rodr{\'{\i}}guez-Ardila}, {Santos}, {Bonatto}, \&
  {Ducati}}]{2011MNRAS.410.2714R}
{Riffel}, R., {Ruschel-Dutra}, D., {Pastoriza}, M.~G., {et~al.}
  2011{\natexlab{b}}, \mnras, 410, 2714

\bibitem[{{Riffel} {et~al.}(2015{\natexlab{b}}){Riffel}, {Ho}, {Mason},
  {Rodr{\'{\i}}guez-Ardila}, {Martins}, {Riffel}, {Diaz}, {Colina},
  {Alonso-Herrero}, {Flohic}, {Gonzalez Martin}, {Lira}, {McDermid}, {Ramos
  Almeida}, {Schiavon}, {Thanjavur}, {Ruschel-Dutra}, {Winge}, \&
  {Perlman}}]{2015MNRAS.446.2823R}
{Riffel}, R.~A., {Ho}, L.~C., {Mason}, R., {et~al.} 2015{\natexlab{b}}, \mnras,
  446, 2823

\bibitem[{{R{\"o}ck} {et~al.}(2015){R{\"o}ck}, {Vazdekis}, {Peletier},
  {Knapen}, \& {Falc{\'o}n-Barroso}}]{2015MNRAS.449.2853R}
{R{\"o}ck}, B., {Vazdekis}, A., {Peletier}, R.~F., {Knapen}, J.~H., \&
  {Falc{\'o}n-Barroso}, J. 2015, \mnras, 449, 2853

\bibitem[{{R{\"o}ck} {et~al.}(2016){R{\"o}ck}, {Vazdekis}, {Ricciardelli},
  {Peletier}, {Knapen}, \& {Falc{\'o}n-Barroso}}]{2016A&A...589A..73R}
{R{\"o}ck}, B., {Vazdekis}, A., {Ricciardelli}, E., {et~al.} 2016, \aap, 589,
  A73

\bibitem[{{Santos} {et~al.}(2002){Santos}, {Alloin}, {Bica}, \&
  {Bonatto}}]{2002IAUS..207....1S}
{Santos}, J.~F.~C.~J., {Alloin}, D., {Bica}, E., \& {Bonatto}, C. 2002, in IAU
  Symposium, Vol. 207, Extragalactic Star Clusters, ed. D.~P. {Geisler}, E.~K.
  {Grebel}, \& D.~{Minniti}, 1--2

\bibitem[{{Sarzi} {et~al.}(2006){Sarzi}, {Falc{\'o}n-Barroso}, {Davies},
  {Bacon}, {Bureau}, {Cappellari}, {de Zeeuw}, {Emsellem}, {Fathi},
  {Krajnovi{\'c}}, {Kuntschner}, {McDermid}, \& {Peletier}}]{sarzetal06}
{Sarzi}, M., {Falc{\'o}n-Barroso}, J., {Davies}, R.~L., {et~al.} 2006, \mnras,
  366, 1151

\bibitem[{{Schechter}(1983)}]{1983ApJS...52..425S}
{Schechter}, P.~L. 1983, \apjs, 52, 425

\bibitem[{{Sharples} {et~al.}(2013){Sharples}, {Bender}, {Agudo Berbel},
  {Bezawada}, {Castillo}, {Cirasuolo}, {Davidson}, {Davies}, {Dubbeldam},
  {Fairley}, {Finger}, {F{\"o}rster Schreiber}, {Gonte}, {Hess}, {Jung},
  {Lewis}, {Lizon}, {Muschielok}, {Pasquini}, {Pirard}, {Popovic}, {Ramsay},
  {Rees}, {Richter}, {Riquelme}, {Rodrigues}, {Saviane}, {Schlichter},
  {Schmidtobreick}, {Segovia}, {Smette}, {Szeifert}, {van Kesteren}, {Wegner},
  \& {Wiezorrek}}]{2013Msngr.151...21S}
{Sharples}, R., {Bender}, R., {Agudo Berbel}, A., {et~al.} 2013, The Messenger,
  151, 21

\bibitem[{{Silva} {et~al.}(2008){Silva}, {Kuntschner}, \&
  {Lyubenova}}]{2008ApJ...674..194S}
{Silva}, D.~R., {Kuntschner}, H., \& {Lyubenova}, M. 2008, \apj, 674, 194

\bibitem[{{Skrutskie} {et~al.}(2006){Skrutskie}, {Cutri}, {Stiening},
  {Weinberg}, {Schneider}, {Carpenter}, {Beichman}, {Capps}, {Chester},
  {Elias}, {Huchra}, {Liebert}, {Lonsdale}, {Monet}, {Price}, {Seitzer},
  {Jarrett}, {Kirkpatrick}, {Gizis}, {Howard}, {Evans}, {Fowler}, {Fullmer},
  {Hurt}, {Light}, {Kopan}, {Marsh}, {McCallon}, {Tam}, {Van Dyk}, \&
  {Wheelock}}]{2006AJ....131.1163S}
{Skrutskie}, M.~F., {Cutri}, R.~M., {Stiening}, R., {et~al.} 2006, \aj, 131,
  1163

\bibitem[{{Smette} {et~al.}(2015){Smette}, {Sana}, {Noll}, {Horst}, {Kausch},
  {Kimeswenger}, {Barden}, {Szyszka}, {Jones}, {Gallenne}, {Vinther},
  {Ballester}, \& {Taylor}}]{2015A&A...576A..77S}
{Smette}, A., {Sana}, H., {Noll}, S., {et~al.} 2015, \aap, 576, A77

\bibitem[{{Smith} {et~al.}(2015){Smith}, {Alton}, {Lucey}, {Conroy}, \&
  {Carter}}]{2015MNRAS.454L..71S}
{Smith}, R.~J., {Alton}, P., {Lucey}, J.~R., {Conroy}, C., \& {Carter}, D.
  2015, \mnras, 454, L71

\bibitem[{{Smith} {et~al.}(2000){Smith}, {Lucey}, {Hudson}, {Schlegel}, \&
  {Davies}}]{2000MNRAS.313..469S}
{Smith}, R.~J., {Lucey}, J.~R., {Hudson}, M.~J., {Schlegel}, D.~J., \&
  {Davies}, R.~L. 2000, \mnras, 313, 469

\bibitem[{{Tamblyn} {et~al.}(1996){Tamblyn}, {Rieke}, {Hanson}, {Close},
  {McCarthy}, \& {Rieke}}]{1996ApJ...456..206T}
{Tamblyn}, P., {Rieke}, G.~H., {Hanson}, M.~M., {et~al.} 1996, \apj, 456, 206

\bibitem[{{Terlevich} {et~al.}(1981){Terlevich}, {Davies}, {Faber}, \&
  {Burstein}}]{1981MNRAS.196..381T}
{Terlevich}, R., {Davies}, R.~L., {Faber}, S.~M., \& {Burstein}, D. 1981,
  \mnras, 196, 381

\bibitem[{{Terndrup} {et~al.}(1991){Terndrup}, {Frogel}, \&
  {Whitford}}]{1991ApJ...378..742T}
{Terndrup}, D.~M., {Frogel}, J.~A., \& {Whitford}, A.~E. 1991, \apj, 378, 742

\bibitem[{{Thomas} {et~al.}(2003){Thomas}, {Maraston}, \&
  {Bender}}]{2003MNRAS.339..897T}
{Thomas}, D., {Maraston}, C., \& {Bender}, R. 2003, \mnras, 339, 897

\bibitem[{{Trager} {et~al.}(1998){Trager}, {Worthey}, {Faber}, {Burstein}, \&
  {Gonz{\'a}lez}}]{1998ApJS..116....1T}
{Trager}, S.~C., {Worthey}, G., {Faber}, S.~M., {Burstein}, D., \&
  {Gonz{\'a}lez}, J.~J. 1998, \apjs, 116, 1

\bibitem[{{van Dokkum}(2001)}]{2001PASP..113.1420V}
{van Dokkum}, P.~G. 2001, \pasp, 113, 1420

\bibitem[{{Vanzi} \& {Rieke}(1997)}]{1997ApJ...479..694V}
{Vanzi}, L. \& {Rieke}, G.~H. 1997, \apj, 479, 694

\bibitem[{{Vanzi} {et~al.}(1996){Vanzi}, {Rieke}, {Martin}, \&
  {Shields}}]{1996ApJ...466..150V}
{Vanzi}, L., {Rieke}, G.~H., {Martin}, C.~L., \& {Shields}, J.~C. 1996, \apj,
  466, 150

\bibitem[{{Veilleux} {et~al.}(1997{\natexlab{a}}){Veilleux}, {Goodrich}, \&
  {Hill}}]{1997ApJ...477..631V}
{Veilleux}, S., {Goodrich}, R.~W., \& {Hill}, G.~J. 1997{\natexlab{a}}, \apj,
  477, 631

\bibitem[{{Veilleux} {et~al.}(1997{\natexlab{b}}){Veilleux}, {Sanders}, \&
  {Kim}}]{1997ApJ...484...92V}
{Veilleux}, S., {Sanders}, D.~B., \& {Kim}, D.-C. 1997{\natexlab{b}}, \apj,
  484, 92

\bibitem[{{Veilleux} {et~al.}(1999){Veilleux}, {Sanders}, \&
  {Kim}}]{1999ApJ...522..139V}
{Veilleux}, S., {Sanders}, D.~B., \& {Kim}, D.-C. 1999, \apj, 522, 139

\bibitem[{{Vernet} {et~al.}(2011){Vernet}, {Dekker}, {D'Odorico}, {Kaper},
  {Kjaergaard}, {Hammer}, {Randich}, {Zerbi}, {Groot}, {Hjorth}, {Guinouard},
  {Navarro}, {Adolfse}, {Albers}, {Amans}, {Andersen}, {Andersen}, {Binetruy},
  {Bristow}, {Castillo}, {Chemla}, {Christensen}, {Conconi}, {Conzelmann},
  {Dam}, {de Caprio}, {de Ugarte Postigo}, {Delabre}, {di Marcantonio},
  {Downing}, {Elswijk}, {Finger}, {Fischer}, {Flores}, {Fran{\c c}ois},
  {Goldoni}, {Guglielmi}, {Haigron}, {Hanenburg}, {Hendriks}, {Horrobin},
  {Horville}, {Jessen}, {Kerber}, {Kern}, {Kiekebusch}, {Kleszcz}, {Klougart},
  {Kragt}, {Larsen}, {Lizon}, {Lucuix}, {Mainieri}, {Manuputy}, {Martayan},
  {Mason}, {Mazzoleni}, {Michaelsen}, {Modigliani}, {Moehler}, {M{\o}ller},
  {Norup S{\o}rensen}, {N{\o}rregaard}, {P{\'e}roux}, {Patat}, {Pena}, {Pragt},
  {Reinero}, {Rigal}, {Riva}, {Roelfsema}, {Royer}, {Sacco}, {Santin},
  {Schoenmaker}, {Spano}, {Sweers}, {Ter Horst}, {Tintori}, {Tromp}, {van
  Dael}, {van der Vliet}, {Venema}, {Vidali}, {Vinther}, {Vola}, {Winters},
  {Wistisen}, {Wulterkens}, \& {Zacchei}}]{2011A&A...536A.105V}
{Vernet}, J., {Dekker}, H., {D'Odorico}, S., {et~al.} 2011, \aap, 536, A105

\bibitem[{{Villaume} {et~al.}(2015){Villaume}, {Conroy}, \&
  {Johnson}}]{2015ApJ...806...82V}
{Villaume}, A., {Conroy}, C., \& {Johnson}, B.~D. 2015, \apj, 806, 82

\bibitem[{{Wallace} \& {Hinkle}(1997)}]{1997ApJS..111..445W}
{Wallace}, L. \& {Hinkle}, K. 1997, \apjs, 111, 445

\bibitem[{{Wallace} \& {Hinkle}(2002)}]{2002AJ....124.3393W}
{Wallace}, L. \& {Hinkle}, K. 2002, \aj, 124, 3393

\bibitem[{{Wallace} {et~al.}(2000){Wallace}, {Meyer}, {Hinkle}, \&
  {Edwards}}]{2000ApJ...535..325W}
{Wallace}, L., {Meyer}, M.~R., {Hinkle}, K., \& {Edwards}, S. 2000, \apj, 535,
  325

\bibitem[{{Wegner} {et~al.}(2003){Wegner}, {Bernardi}, {Willmer}, {da Costa},
  {Alonso}, {Pellegrini}, {Maia}, {Chaves}, \&
  {Rit{\'e}}}]{2003AJ....126.2268W}
{Wegner}, G., {Bernardi}, M., {Willmer}, C.~N.~A., {et~al.} 2003, \aj, 126,
  2268

\bibitem[{{Wegner} {et~al.}(1999){Wegner}, {Colless}, {Saglia}, {McMahan},
  {Davies}, {Burstein}, \& {Baggley}}]{1999MNRAS.305..259W}
{Wegner}, G., {Colless}, M., {Saglia}, R.~P., {et~al.} 1999, \mnras, 305, 259

\bibitem[{{Worthey}(1994)}]{1994ApJS...95..107W}
{Worthey}, G. 1994, \apjs, 95, 107

\bibitem[{{Worthey} {et~al.}(1992){Worthey}, {Faber}, \&
  {Gonzalez}}]{1992ApJ...398...69W}
{Worthey}, G., {Faber}, S.~M., \& {Gonzalez}, J.~J. 1992, \apj, 398, 69

\bibitem[{{Worthey} {et~al.}(1994){Worthey}, {Faber}, {Gonzalez}, \&
  {Burstein}}]{1994ApJS...94..687W}
{Worthey}, G., {Faber}, S.~M., {Gonzalez}, J.~J., \& {Burstein}, D. 1994,
  \apjs, 94, 687

\bibitem[{{York} {et~al.}(2000){York}, {Adelman}, {Anderson}, {Anderson},
  {Annis}, {Bahcall}, {Bakken}, {Barkhouser}, {Bastian}, {Berman}, {Boroski},
  {Bracker}, {Briegel}, {Briggs}, {Brinkmann}, {Brunner}, {Burles}, {Carey},
  {Carr}, {Castander}, {Chen}, {Colestock}, {Connolly}, {Crocker}, {Csabai},
  {Czarapata}, {Davis}, {Doi}, {Dombeck}, {Eisenstein}, {Ellman}, {Elms},
  {Evans}, {Fan}, {Federwitz}, {Fiscelli}, {Friedman}, {Frieman}, {Fukugita},
  {Gillespie}, {Gunn}, {Gurbani}, {de Haas}, {Haldeman}, {Harris}, {Hayes},
  {Heckman}, {Hennessy}, {Hindsley}, {Holm}, {Holmgren}, {Huang}, {Hull},
  {Husby}, {Ichikawa}, {Ichikawa}, {Ivezi{\'c}}, {Kent}, {Kim}, {Kinney},
  {Klaene}, {Kleinman}, {Kleinman}, {Knapp}, {Korienek}, {Kron}, {Kunszt},
  {Lamb}, {Lee}, {Leger}, {Limmongkol}, {Lindenmeyer}, {Long}, {Loomis},
  {Loveday}, {Lucinio}, {Lupton}, {MacKinnon}, {Mannery}, {Mantsch}, {Margon},
  {McGehee}, {McKay}, {Meiksin}, {Merelli}, {Monet}, {Munn}, {Narayanan},
  {Nash}, {Neilsen}, {Neswold}, {Newberg}, {Nichol}, {Nicinski}, {Nonino},
  {Okada}, {Okamura}, {Ostriker}, {Owen}, {Pauls}, {Peoples}, {Peterson},
  {Petravick}, {Pier}, {Pope}, {Pordes}, {Prosapio}, {Rechenmacher}, {Quinn},
  {Richards}, {Richmond}, {Rivetta}, {Rockosi}, {Ruthmansdorfer}, {Sandford},
  {Schlegel}, {Schneider}, {Sekiguchi}, {Sergey}, {Shimasaku}, {Siegmund},
  {Smee}, {Smith}, {Snedden}, {Stone}, {Stoughton}, {Strauss}, {Stubbs},
  {SubbaRao}, {Szalay}, {Szapudi}, {Szokoly}, {Thakar}, {Tremonti}, {Tucker},
  {Uomoto}, {Vanden Berk}, {Vogeley}, {Waddell}, {Wang}, {Watanabe},
  {Weinberg}, {Yanny}, {Yasuda}, \& {SDSS Collaboration}}]{2000AJ....120.1579Y}
{York}, D.~G., {Adelman}, J., {Anderson}, Jr., J.~E., {et~al.} 2000, \aj, 120,
  1579

\bibitem[{{Zamora} {et~al.}(2015){Zamora}, {Garc{\'{\i}}a-Hern{\'a}ndez},
  {Allende Prieto}, {Carrera}, {Koesterke}, {Edvardsson}, {Castelli}, {Plez},
  {Bizyaev}, {Cunha}, {Garc{\'{\i}}a P{\'e}rez}, {Gustafsson}, {Holtzman},
  {Lawler}, {Majewski}, {Manchado}, {M{\'e}sz{\'a}ros}, {Shane}, {Shetrone},
  {Smith}, \& {Zasowski}}]{2015AJ....149..181Z}
{Zamora}, O., {Garc{\'{\i}}a-Hern{\'a}ndez}, D.~A., {Allende Prieto}, C.,
  {et~al.} 2015, \aj, 149, 181

\bibitem[{{Zibetti} {et~al.}(2013){Zibetti}, {Gallazzi}, {Charlot}, {Pierini},
  \& {Pasquali}}]{2013MNRAS.428.1479Z}
{Zibetti}, S., {Gallazzi}, A., {Charlot}, S., {Pierini}, D., \& {Pasquali}, A.
  2013, \mnras, 428, 1479

\end{thebibliography}

\clearpage
\begin{appendix}

\section{Literature data for the sample galaxies}

\begin{table*}
\caption{Previous measurements of the heliocentric velocities $V_{\rm H}$ for 
the sample galaxies.}
\begin{center}
\begin{small}
\begin{tabular}{@{ }l@{ }c@{ }c@{ }c@{ }c@{ }c@{ }c@{ }c@{ }c@{ }c@{ }}
\hline
\noalign{\smallskip}
\multicolumn{1}{c}{Galaxy} &
\multicolumn{9}{c}{$V_{\rm H}$\,[\kms]} \\ 
\noalign{\smallskip}
\multicolumn{1}{c}{ID} &
\multicolumn{1}{c}{T81} &
\multicolumn{1}{c}{S83} &
\multicolumn{1}{c}{D87/F89} &
\multicolumn{1}{c}{W99/C01} &
\multicolumn{1}{c}{S00} & 
\multicolumn{1}{c}{B02} &
\multicolumn{1}{c}{W03} &
\multicolumn{1}{c}{H07} &
\multicolumn{1}{c}{H12} \\
\noalign{\smallskip}
\hline
\noalign{\smallskip}  
NGC\,0584 & 1968 &  ... & 1875 & 1854 & 1802 & ...        & ...        &  ... & 1872$\pm$45 \\
NGC\,0636 &  ... &  ... & 1805 &  ... &  ... & ...        & ...        &  ... & 2043$\pm$45 \\
NGC\,0897 &  ... & 2008 &  ... &  ... &  ... & ...        & ...        & 2026 & 4666$\pm$45 \\
NGC\,1357 &  ... &  ... &  ... &  ... &  ... & ...        & ...        &  ... & 1965$\pm$45 \\
NGC\,1425 &  ... &  ... &  ... &  ... &  ... & ...        & ...        &  ... & 1564$\pm$45 \\
NGC\,1700 & 3911 &  ... & 3881 &  ... & 3908 & ...        & ...        &  ... & 3891$\pm$45 \\
NGC\,2613 &  ... &  ... &  ... &  ... &  ... & ...        & ...        &  ... & 1693$\pm$45 \\
NGC\,3115 &  ... &  ... &  698 &  732 &  726 & ...        & 668$\pm$10 &  657 &  663$\pm$4  \\
NGC\,3377 &  741 &  ... &  689 &  643 &  678 & ...        & ...        &  ... &  665$\pm$2  \\
NGC\,3379 &  741 &  ... &  922 &  908 &  915 & ...        & 918$\pm$15 &  910 &  911$\pm$2  \\
NGC\,3423 &  ... &  ... &  ... &  ... &  ... & ...        & ...        & 1012 & 1000$\pm$3  \\
NGC\,4415 &  ... &  ... &  ... &  ... &  ... & 933$\pm$19 & 933$\pm$33 &  ... &  907$\pm$2  \\
NGC\,7424 &  ... &  ... &  ... &  ... &  ... & ...        & ...        &  940 &  945$\pm$45 \\
\noalign{\smallskip}
\hline
\noalign{\medskip}
\end{tabular}
\end{small}
\label{tab:lit_vh}
\tablebib{
B02 - \citet{2002AJ....123.2990B};
C01 - \citet{2001MNRAS.328.1039C};
D87 - \citet{1987ApJS...64..581D};
F89 - \citet{1989ApJS...69..763F};
H07 - \citet{2007ApJ...668...94H};
H12 - \citet{2012ApJS..199...26H};
S83 - \citet{1983ApJS...52..425S};
S00 - \citet{2000MNRAS.313..469S};
T98 - \citet{1998ApJS..116....1T};
W99 - \citet{1999MNRAS.305..259W};
W03 - \citet{2003AJ....126.2268W}.
}
\end{center}
\end{table*}

\begin{table*}
\caption{Previous measurements of the velocity dispersions $\sigma_V$ for the 
sample galaxies.}
\begin{center}
\begin{small}
\begin{tabular}{@{ }l@{ }c@{ }c@{ }c@{ }c@{ }c@{ }c@{ }c@{ }c@{ }c@{ }c@{ }c@{ }c@{ }c@{ }}
\hline
\noalign{\smallskip}
\multicolumn{1}{c}{Galaxy} &
\multicolumn{13}{c}{$\sigma_V$\,[\kms]} \\ 
\noalign{\smallskip}
\multicolumn{1}{c}{ID} &
\multicolumn{1}{c}{T81} &
\multicolumn{1}{c}{S83} &
\multicolumn{1}{c}{D87/F89} &
\multicolumn{1}{c}{B89} &
\multicolumn{1}{c}{Fr89} &
\multicolumn{1}{c}{Mc95} &
\multicolumn{1}{c}{T98} &
\multicolumn{1}{c}{W99/C01} &
\multicolumn{1}{c}{S00} & 
\multicolumn{1}{c}{B02} & 
\multicolumn{1}{c}{W03} & 
\multicolumn{1}{c}{H07} &
\multicolumn{1}{c}{H09} \\
\noalign{\smallskip}
\hline
\noalign{\smallskip}  
NGC\,0584 & 225 & ...        & 217 & ...        & ...       & 230 & 198$\pm$1  & 179$\pm$6  & 210$\pm$4 & ...          & ...        & ...            & ...           \\
NGC\,0636 & ... & ...        & 156 & ...        & 151$\pm$5 & 166 & 162$\pm$1  & ...        & ...       & ...          & ...        & ...            & ...           \\
NGC\,0897 & ... & ...        & ... & ...        & ...       & ... & ...        & ...        & ...       & ...          & ...        & 124.0$\pm$14.4 & ...           \\
NGC\,1357 & ... & 121$\pm$14 & ... & ...        & ...       & 119 & ...        & ...        & ...       & ...          & ...        & ...            & ...           \\
NGC\,1425 & ... & ...        & ... & 139$\pm$15 & ...       & ... & ...        & ...        & ...       & ...          & ...        & ...            & ...           \\
NGC\,1700 & 260 & ...        & 233 & ...        & 240$\pm$5 & 243 & 226$\pm$1  & ...        & 228$\pm$7 & ...          & ...        & ...            & ...           \\
NGC\,2613 & ... & ...        & ... & ...        & ...       & 177 & ...        & ...        & ...       & ...          & ...        & ...            & ...           \\
NGC\,3115 & ... & ...        & 266 & ...        & ...       & 264 & 280$\pm$10 & 239$\pm$14 & 259$\pm$4 & ...          & 282$\pm$9  & 252.1$\pm$5.0  & 258.5$\pm$9.3 \\
NGC\,3377 & 176 & ...        & 131 & ...        & ...       & 148 & 126$\pm$2  & 116$\pm$15 & 141$\pm$3 & ...          & ...        & ...            & 156.9$\pm$7.6 \\
NGC\,3379 & 241 & ...        & 201 & ...        & 189$\pm$3 & 209 & 212$\pm$1  & 185$\pm$7  & 214$\pm$4 & ...          & 222$\pm$10 & 207.1$\pm$2.2  & 207.1$\pm$8.4 \\
NGC\,3423 & ... & ...        & ... & ...        & ...       & ... & ...        & ...        & ...       & ...          & ...        &  30.4$\pm$4.6  &  54.6$\pm$8.5 \\
NGC\,4415 & ... & ...        & ... & ...        & ...       & ... & ...        & ...        & ...       & 47.4$\pm$5.4 &  48$\pm$6  & ...            & ...           \\
NGC\,7424 & ... & ...        & ... & ...        & ...       & ... & ...        & ...        & ...       & ...          & ...        &  15.6$\pm$2.3  & ...           \\
\noalign{\smallskip}
\hline
\noalign{\medskip}
\end{tabular}
\end{small}
\label{tab:lit_sigma}
\tablebib{
B02  - \citet{2002AJ....123.2990B};
B89  - \citet{1989A&A...221..236B};
C01  - \citet{2001MNRAS.328.1039C};
D87  - \citet{1987ApJS...64..581D};
F89  - \citet{1989ApJS...69..763F};
Fr89 - \citet{1989ApJ...344..613F};
H07  - \citet{2007ApJ...668...94H};
H09  - \citet{2009ApJS..183....1H};
Mc95 - \citet{1995ApJS..100..105M};
S83  - \citet{1983ApJS...52..425S};
S00  - \citet{2000MNRAS.313..469S};
T98  - \citet{1998ApJS..116....1T};
W99  - \citet{1999MNRAS.305..259W};
W03  - \citet{2003AJ....126.2268W}.
}
\end{center}
\end{table*}

\begin{table*}
\caption{Previous measurements of the \Hb\ and \Mgd\  line-strength indices for the
sample galaxies.}
\begin{center}
\begin{small}
\begin{tabular}{@{ }l@{ }c@{ }c@{ }c@{ }c@{ }c@{ }c@{ }c@{ }c@{ }c@{ }c@{ }c@{ }c@{ }c@{ }c@{ }}
\hline
\noalign{\smallskip}
\multicolumn{1}{c}{Galaxy} &
\multicolumn{1}{c}{\Hb\,[\AA]} &
\multicolumn{8}{c}{\Mgd\,[mag]} \\
\noalign{\smallskip}
\multicolumn{1}{c}{ID} &
\multicolumn{1}{c}{T98} &
\multicolumn{1}{c}{T81} & 
\multicolumn{1}{c}{D87/F89} & 
\multicolumn{1}{c}{W92} &
\multicolumn{1}{c}{T98} &
\multicolumn{1}{c}{W99/C01} &
\multicolumn{1}{c}{S00} &
\multicolumn{1}{c}{B02} &
\multicolumn{1}{c}{W03} \\
\noalign{\smallskip}
\hline
\noalign{\smallskip}  
NGC\,0584 & 1.537$\pm$0.121 & 0.266 & 0.283 & 0.295 & 0.2918$\pm$0.0044 & 0.337$\pm$0.0088 & 0.278$\pm$0.005 & ...             & ...             \\
NGC\,0636 & 1.963$\pm$0.206 & ...   & 0.273 & 0.285 & 0.2846$\pm$0.0074 & ...              & ...             & ...             & ...             \\
NGC\,1700~&~2.023$\pm$0.209~& 0.267 & 0.278 & 0.296 & 0.2768$\pm$0.0073 & ...              & 0.283$\pm$0.004 & ...             & ...             \\
NGC\,3115 & 1.268$\pm$0.150 & ...   & 0.309 & 0.336 & 0.3380$\pm$0.0055 & ...              & 0.326$\pm$0.003 & ...             & 0.291$\pm$0.017 \\
NGC\,3377 & 1.722$\pm$0.105 & 0.287 & 0.270 & 0.287 & 0.2903$\pm$0.0038 &~0.289$\pm$0.0177~& 0.267$\pm$0.004 & ...             & ...             \\
NGC\,3379 & 1.243$\pm$0.168 & 0.329 & 0.308 & 0.341 & 0.3371$\pm$0.0061 & ...              & 0.314$\pm$0.004 & ...             & 0.301$\pm$0.021 \\
NGC\,4415 & ...               & ...   & ...   & ...   & ...               & ...              & ...             &~0.131$\pm$0.009~& 0.129$\pm$0.021 \\
\noalign{\smallskip}
\hline
\noalign{\medskip}
\end{tabular}
\end{small}
\label{tab:lit_indices1}
\tablebib{
B02 - \citet{2002AJ....123.2990B};
C01 - \citet{2001MNRAS.328.1039C};
D87 - \citet{1987ApJS...64..581D};
F89 - \citet{1989ApJS...69..763F};
S00 - \citet{2000MNRAS.313..469S};
T81 - \citet{1981MNRAS.196..381T};
T98 - \citet{1998ApJS..116....1T};
W03 - \citet{2003AJ....126.2268W};
W92 - \citet{1992ApJ...398...69W};
W99 - \citet{1999MNRAS.305..259W}.
}
\end{center}
\end{table*}

\begin{table*}
\caption{Previous measurements of the spectral indices \Mgb, 
\Mgb\protect{$^\prime$}, Fe5270 and Fe5335 for sample galaxies.}
\begin{center}
\begin{small}
\begin{tabular}{@{ }l@{ }c@{ }c@{ }c@{ }c@{ }c@{ }c@{ }c@{ }}
\hline
\noalign{\smallskip}
\multicolumn{1}{c}{Galaxy} &
\multicolumn{2}{c}{\Mgb\,[\AA]} &
\multicolumn{1}{c}{\Mgb\protect{$^\prime$}\,[mag]} &
\multicolumn{2}{c}{Fe5270\,[\AA]} &
\multicolumn{2}{c}{Fe5335\,[\AA]} \\
\noalign{\smallskip}
\multicolumn{1}{c}{ID} &
\multicolumn{1}{c}{T98} &
\multicolumn{1}{c}{W99/C01} &
\multicolumn{1}{c}{S00} &
\multicolumn{1}{c}{W92} &
\multicolumn{1}{c}{T98} &
\multicolumn{1}{c}{W92} &
\multicolumn{1}{c}{T98} \\
\noalign{\smallskip}
\hline
\noalign{\smallskip}  
NGC\,0584 &~4.458$\pm$0.136 & 4.28$\pm$0.11 & 0.155$\pm$0.005 & 3.27 & 3.259$\pm$0.132 & 3.23 & 3.2582$\pm$0.1681 \\
NGC\,0636 &~4.118$\pm$0.227 & ...           & ...             & 3.32 & 3.314$\pm$0.218 & 3.04 & 2.8043$\pm$0.2653 \\
NGC\,1700 &~3.791$\pm$0.235 & ...           & 0.163$\pm$0.005 & 3.31 & 3.496$\pm$0.229 & 3.06 & 3.1935$\pm$0.2965 \\
NGC\,3115 &~4.688$\pm$0.226 & 4.44$\pm$0.34 & 0.178$\pm$0.004 & 3.73 & 3.371$\pm$0.205 & 3.17 & 3.0822$\pm$0.3187 \\
NGC\,3377 &~4.330$\pm$0.114 & 3.72$\pm$0.40 & 0.155$\pm$0.005 & 2.89 & 2.860$\pm$0.109 & 2.69 & 2.5965$\pm$0.1298 \\
NGC\,3379~&~4.720$\pm$0.192~& 4.76$\pm$0.21 & 0.179$\pm$0.003 & 3.02 & 2.957$\pm$0.184 & 2.89 & 2.8516$\pm$0.2358 \\
\noalign{\smallskip}
\hline
\noalign{\medskip}
\end{tabular}
\end{small}
\label{tab:lit_indices2}
\tablebib{
C01 - \citet{2001MNRAS.328.1039C};
S00 - \citet{2000MNRAS.313..469S};
T98 - \citet{1998ApJS..116....1T};
W92 - \citet{1992ApJ...398...69W};
W99 - \citet{1999MNRAS.305..259W}.
}
\end{center}
\end{table*}

\begin{table}
\caption{List of the available spectra of the sample galaxies.}
\begin{center}
\begin{tabular}{@{}l@{ }c@{}c@{}c@{}l@{}}
\hline
\noalign{\smallskip}
\multicolumn{1}{l}{Galaxy} &
\multicolumn{1}{c}{Spectral Range} &
\multicolumn{1}{c}{Resolu-} &
\multicolumn{1}{c}{Flux} &
\multicolumn{1}{l}{Reference}  \\
\multicolumn{1}{c}{ID} &
\multicolumn{1}{c}{[nm]} &
\multicolumn{1}{c}{tion\,[nm]} &
\multicolumn{1}{c}{calibr.} &
\multicolumn{1}{c}{}  \\
\noalign{\smallskip}
\hline
\noalign{\smallskip}
NGC\,0584 & 363.60-803.60 & 0.40 & yes & L96,\,S02 \\
          & 391.30-757.81 & 0.58 & no  & J09 \\
NGC\,0636 & 391.30-757.81 & 0.58 & no  & J09 \\
NGC\,0897 & 362.38-803.32 & 0.90 & no  & C03 \\
          & 403.68-850.84 & 0.58 & no  & J09 \\
NGC\,1357 & 365.00-710.00 & 0.45 & yes & K92 \\
          & 350.97-746.09 & 0.70 & yes & G04 \\
          & 391.30-757.81 & 0.58 & no  & J09 \\
NGC\,1425 & 363.60-804.60 & 0.40 & yes & L96,\,S02 \\
          & 393.23-759.70 & 0.58 & no  & J09 \\
NGC\,1600 & 363.60-796.20 & 0.40 & yes & L96,\,S02 \\
NGC\,1700 & 391.30-757.81 & 0.58 & no  & J09 \\
NGC\,2613 & 394.16-760.22 & 0.58 & no  & J09 \\
NGC\,3115 & 424.00-507.60 & 0.40 & yes & H95 \\
          & 622.60-684.20 & 0.25 & yes & H95 \\
          & 300.00-564.80 & 0.40 & yes & L96,\,S02 \\
          & 369.00-807.00 & 0.40 & yes & L96,\,S02 \\
          & 636.00-978.60 & 0.40 & yes & L96,\,S02 \\
          & 602.37-812.01 & 0.60 & yes & G13 \\
NGC\,3377 & 422.00-507.20 & 0.40 & yes & H95 \\
          & 619.50-684.10 & 0.25 & yes & H95 \\
NGC\,3379 & 365.00-710.00 & 0.45 & yes & K92 \\
          & 418.46-528.43 & 0.40 & yes & H95 \\
          & 622.15-686.46 & 0.25 & yes & H95 \\
          & 369.00-814.20 & 0.40 & yes & L96,\,S02 \\
          & 360.35-703.08 & 0.70 & yes & G04 \\
NGC\,3423 & 420.83-507.09 & 0.40 & yes & H95 \\
          & 619.27-684.09 & 0.25 & yes & H95 \\
          & 454.50-706.52 & 0.48 & no  & F99 \\
NGC\,7424 & 300.00-521.40 & 0.40 & yes & L96,\,S02 \\
          & 403.68-850.84 & 0.58 & no  & J09 \\
\hline
\noalign{\bigskip}
\label{tab:spectra_literature}
\end{tabular}
\tablebib{
C03 - \citet{2003astro.ph..6581C};
F99 - \citet{1999PASP..111..438F};
G04 - \citet{2004A&A...417..499G};
G13 - \citet{2013A&A...558A..68G};
H95 - \citet{1995ApJS...98..477H};
J09 - \citet{2009MNRAS.399..683J};
K92 - \citet{1992ApJS...79..255K};
L96 - \citet{1996PASP..108..996L};
S02 - \citet{2002IAUS..207....1S}.
}
\end{center}
\end{table}

\begin{figure}
\centering
\includegraphics[angle=0.0,width=10cm]{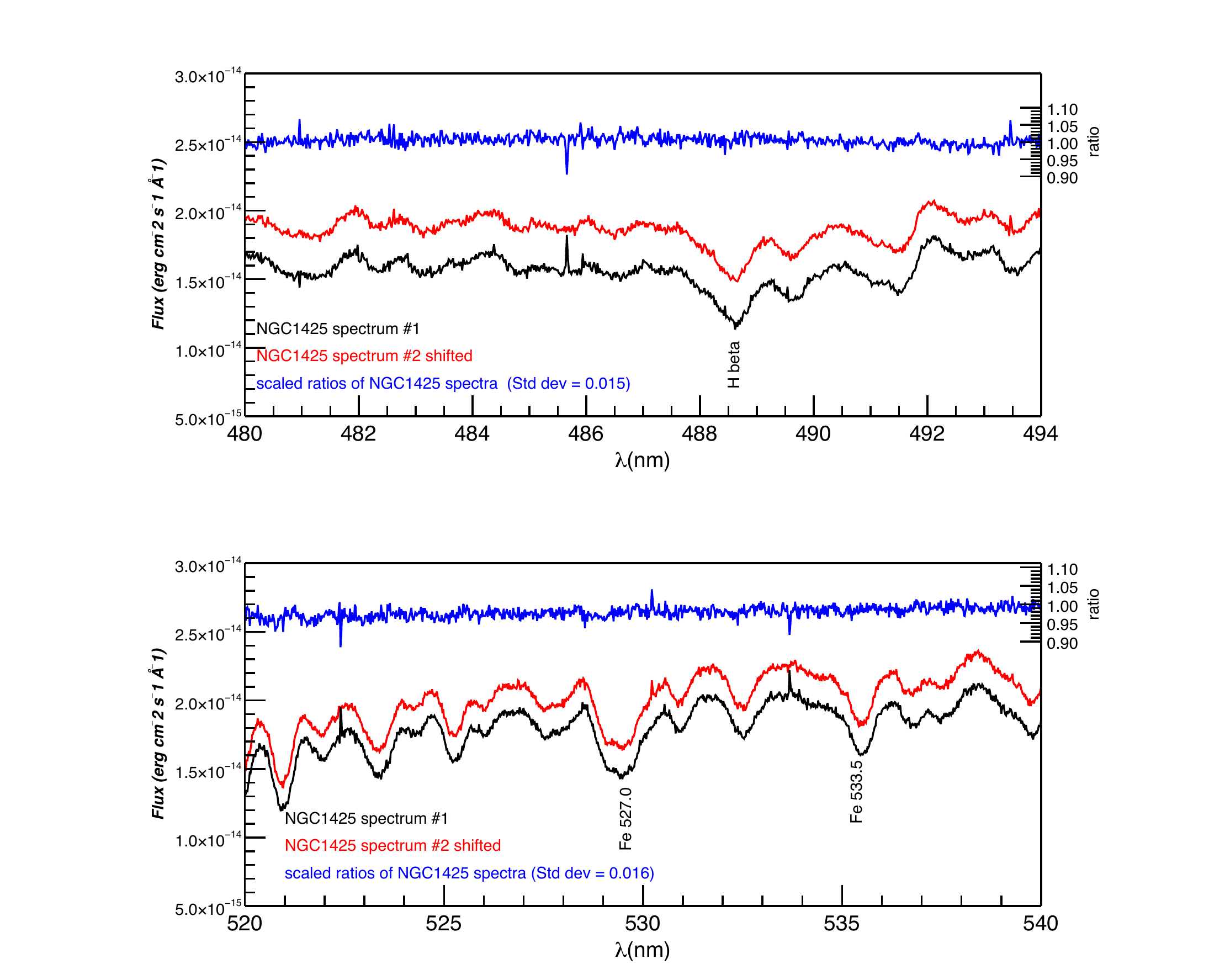}\\
\includegraphics[angle=0.0,width=10cm]{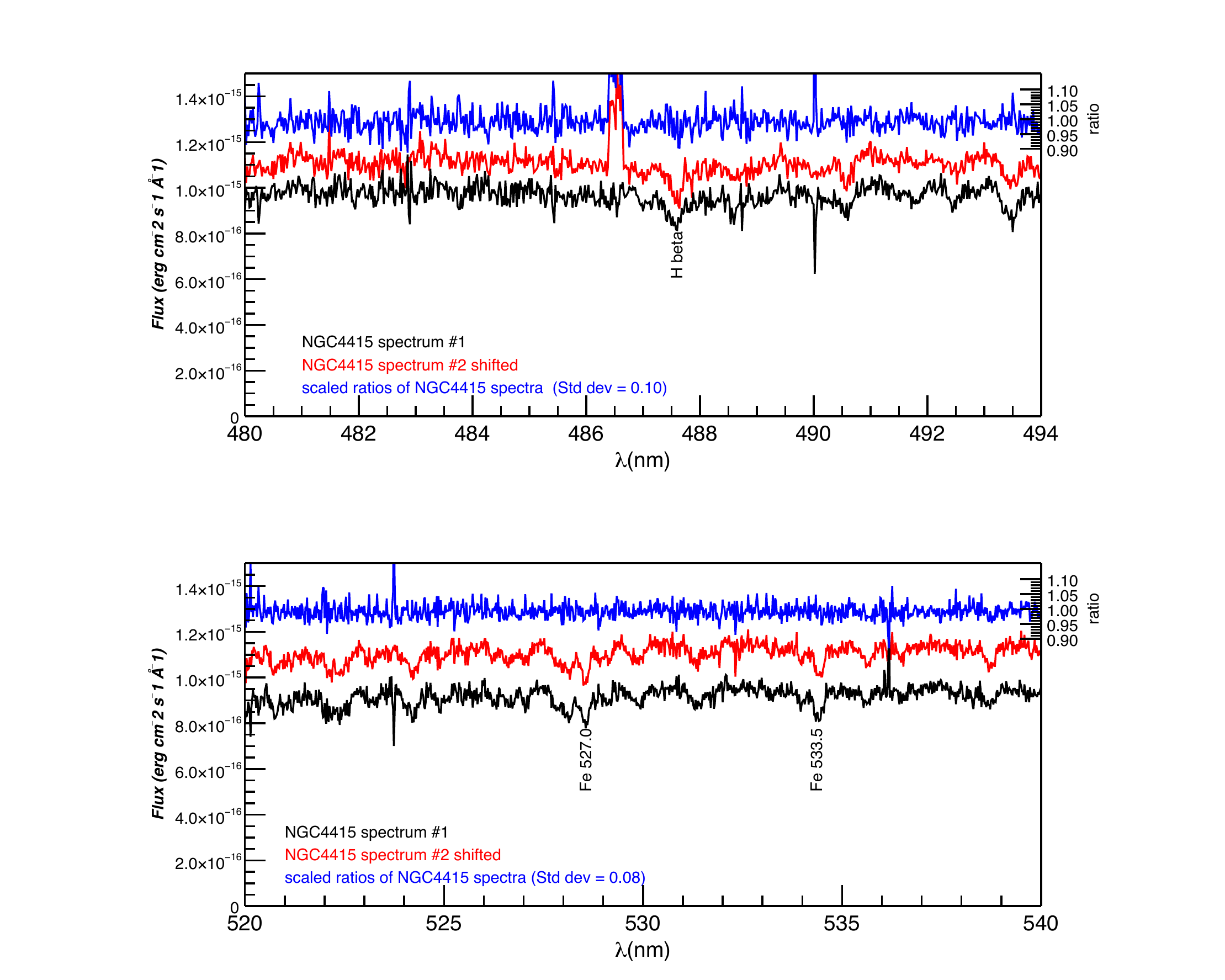}\\
\caption{Comparison of selected regions of the spectra from two 
observations of NGC\,1425 (top two panels) and NGC\,4415 (bottom two panels).}
\label{fig:spectra_comp}
\end{figure}

\begin{figure}
\centering
\includegraphics[angle=0.0,width=8.8cm]{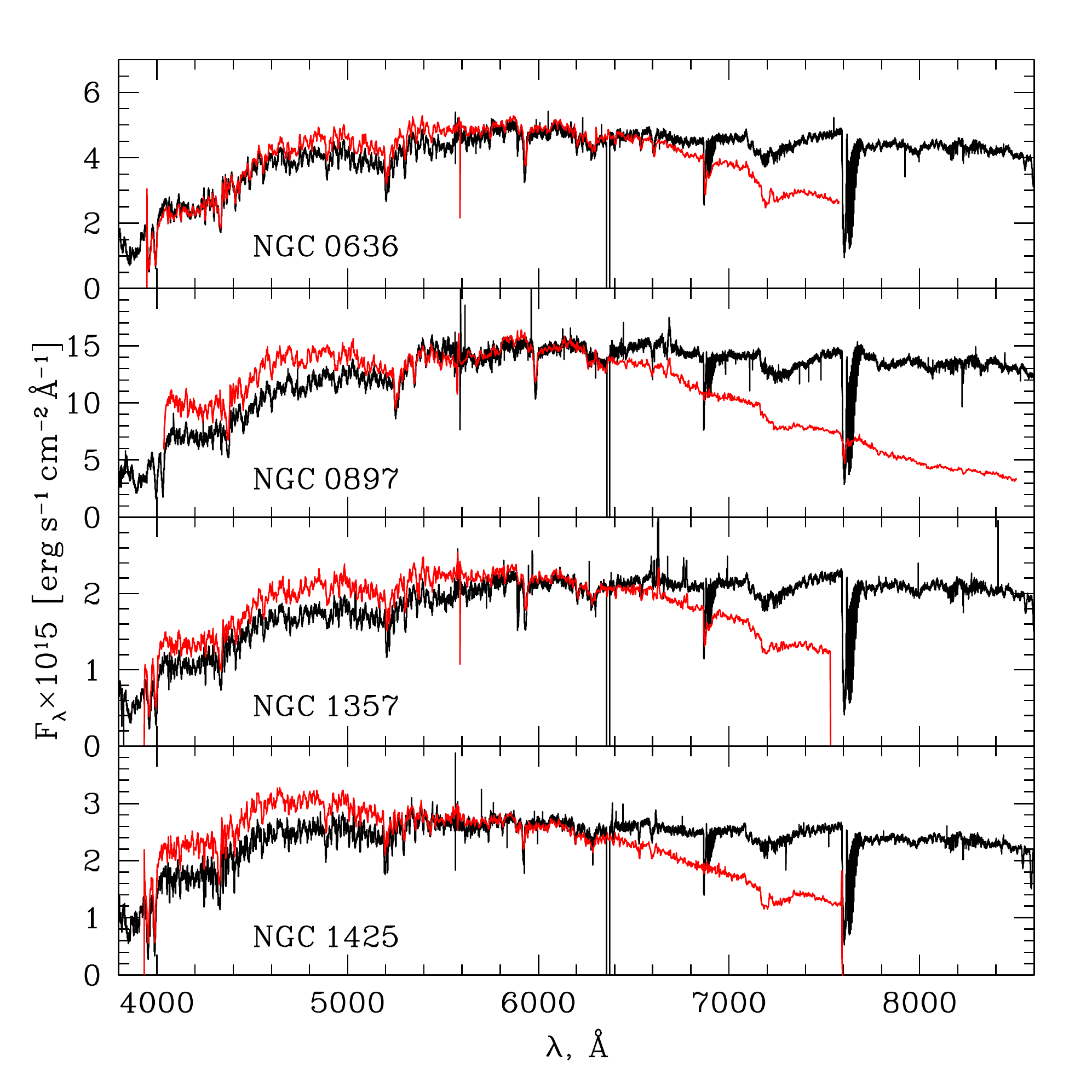}\\
\includegraphics[angle=0.0,width=8.8cm]{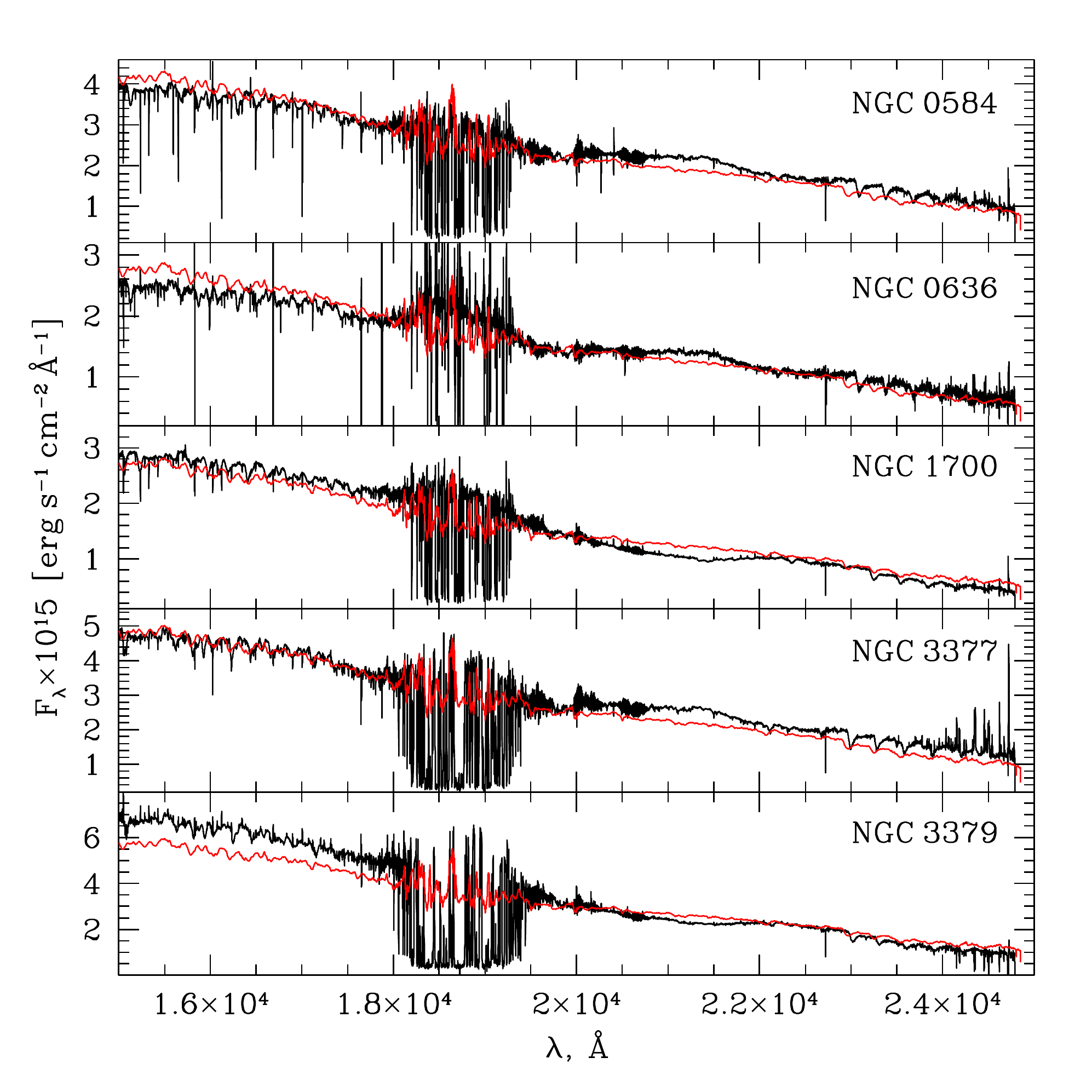}\\
\caption{Further comparison of our spectra (black) with literature (red) spectra 
from \citet{2009MNRAS.399..683J} for some of the galaxies in common between both
sample (top) and from \citet{2009A&A...497...41C} for our bonafide ellipticals 
and their average spectrum of ellipticals. The literature spectra were always 
scaled to match the median flux level of our spectra.}
\label{fig:spectra_comp_J09}
\end{figure}

\section{Spectra of the sample galaxies}

\begin{figure*}
\centering
\includegraphics[angle=0.0,width=6.0cm]{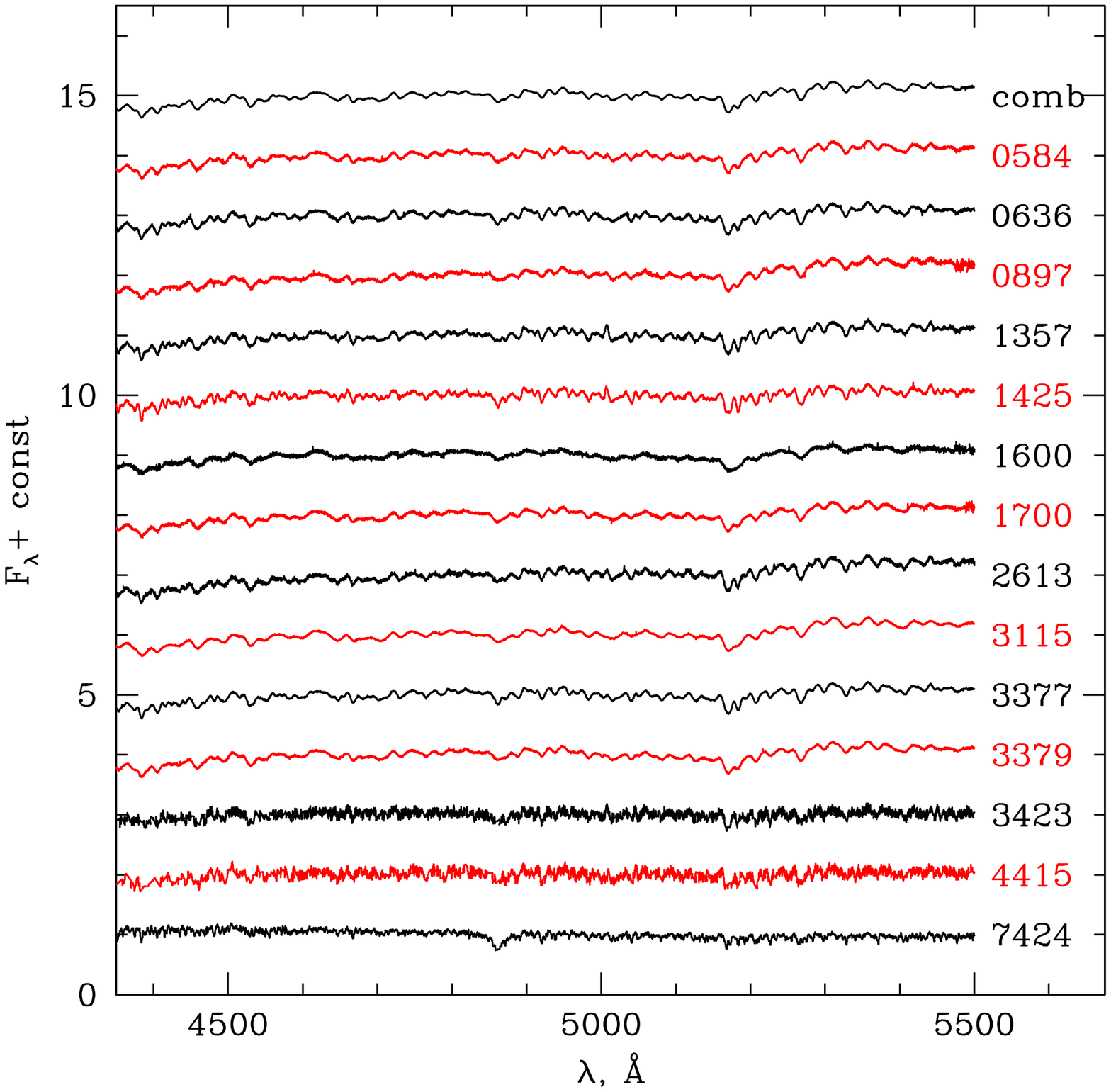}
\includegraphics[angle=0.0,width=6.0cm]{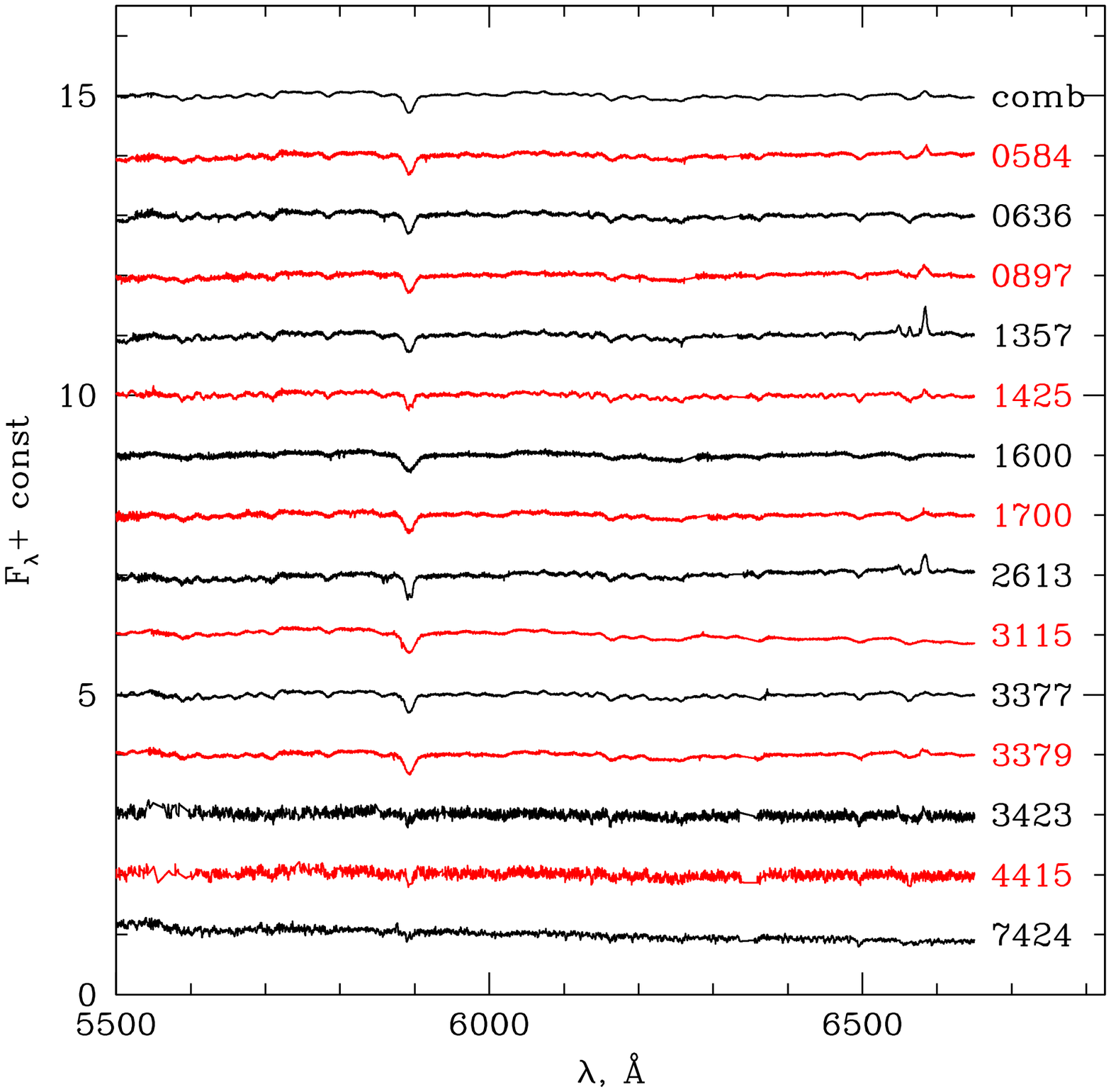}\\
\includegraphics[angle=0.0,width=6.0cm]{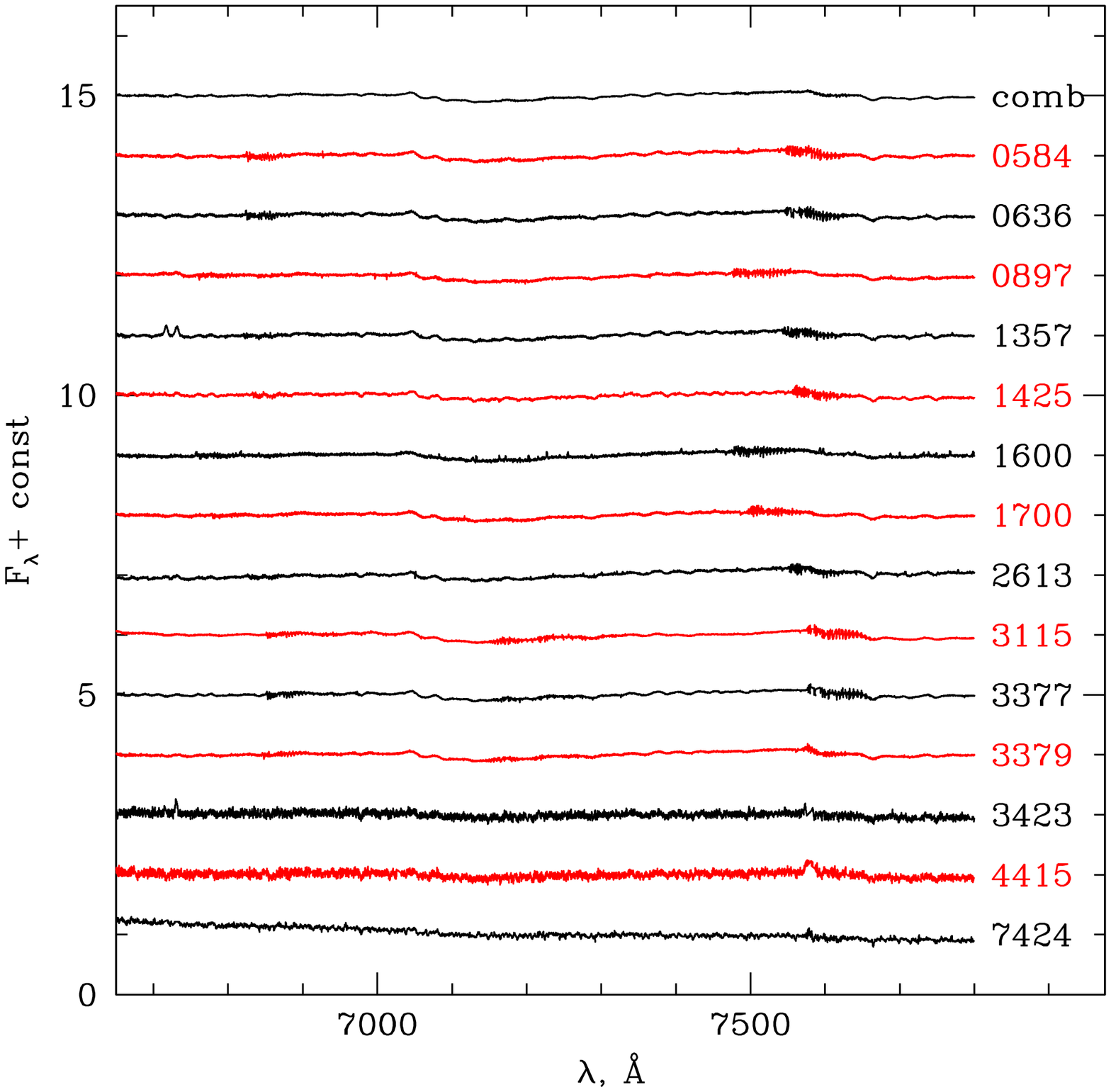}
\includegraphics[angle=0.0,width=6.0cm]{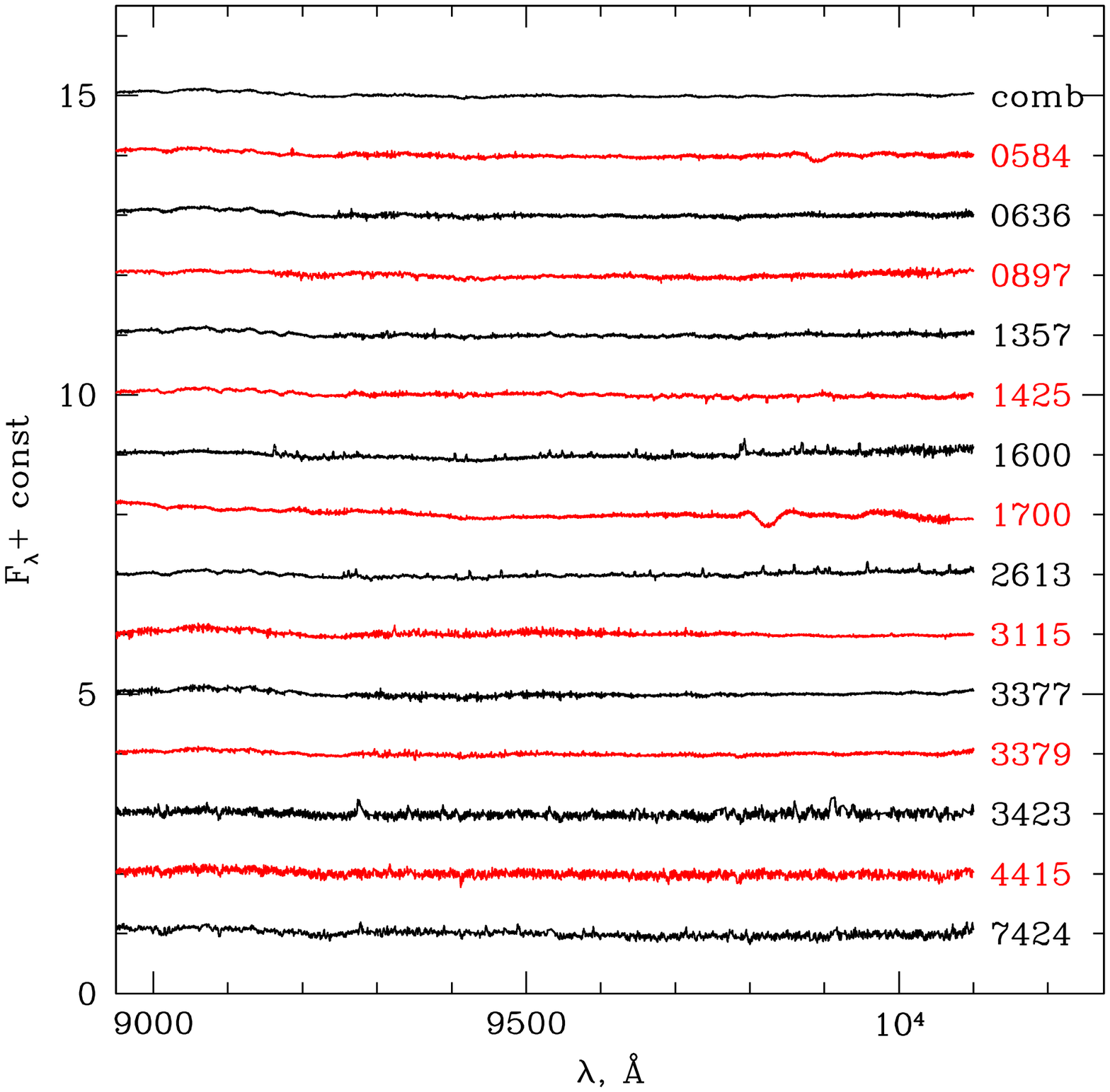}\\
\includegraphics[angle=0.0,width=6.0cm]{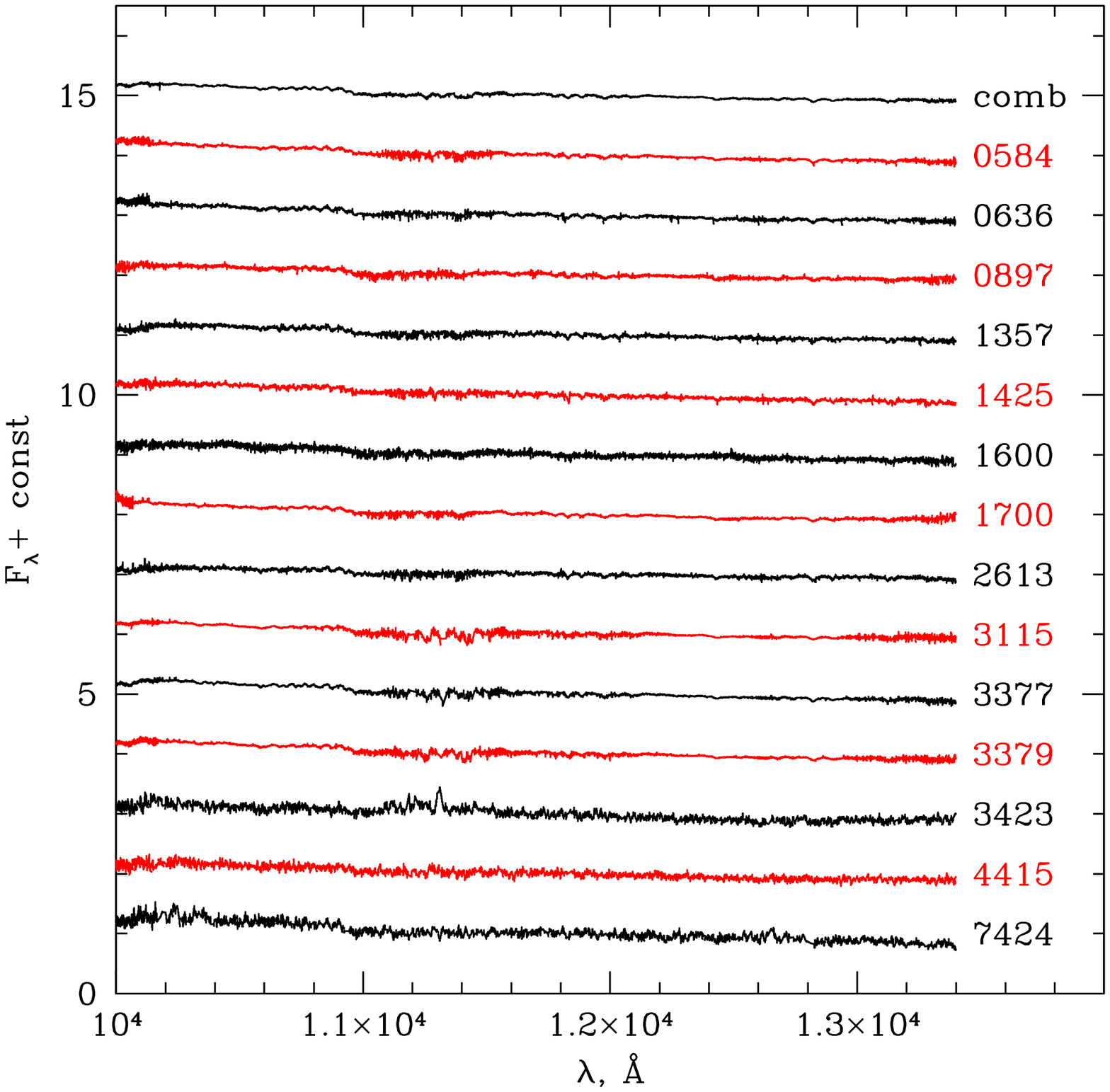}
\includegraphics[angle=0.0,width=6.0cm]{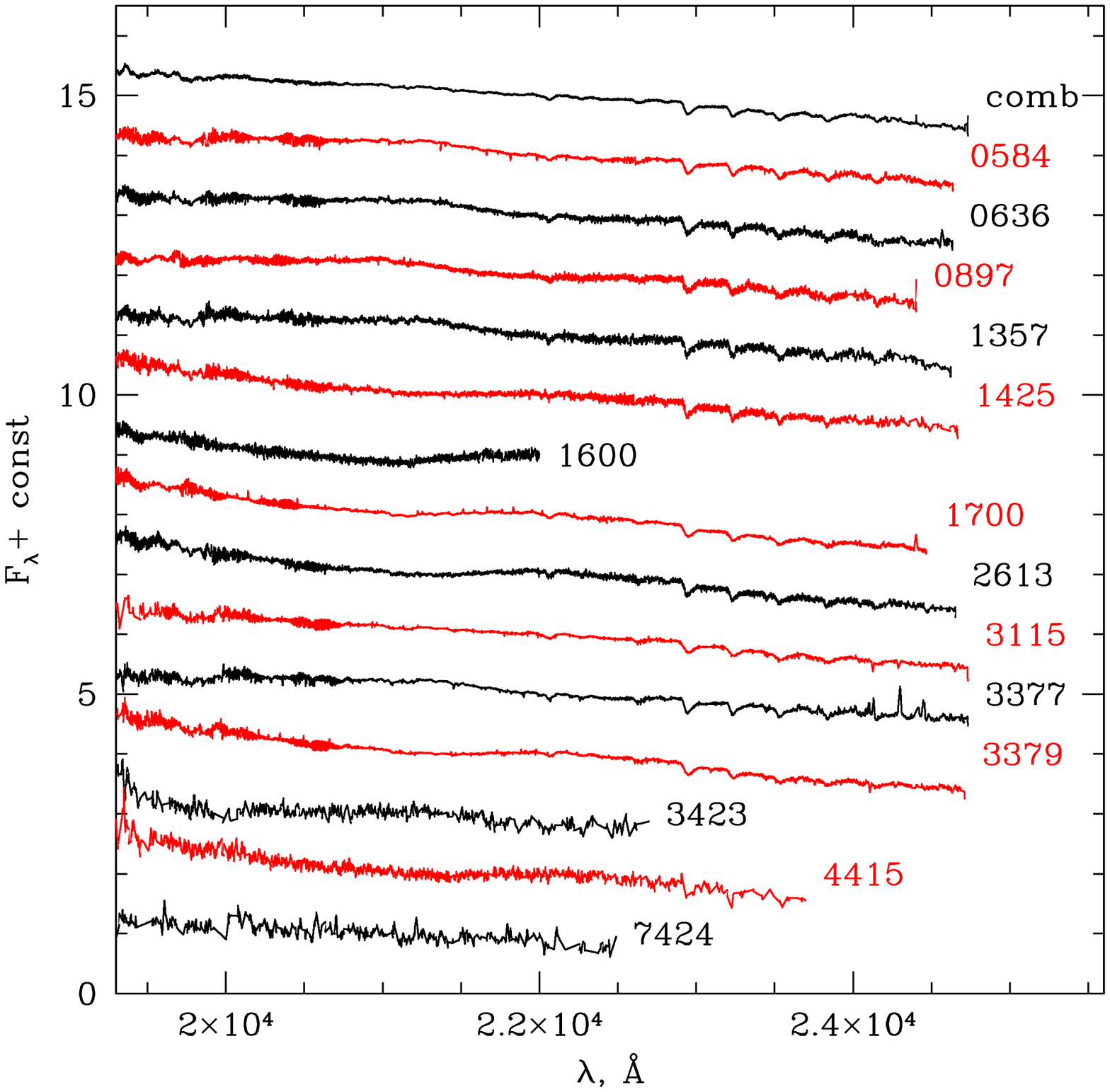}\\
\caption{As in Fig.\,\ref{fig:spectra1}, but for the remaining spectra of our sample galaxies.}
\label{fig:spectra2}
\end{figure*}

\section{Apparent central colors of the sample galaxies}

\begin{table*}
\caption{Apparent colors for the central 1.3\arcsec$\times$4\arcsec 
apertures of the galaxies in our sample, derived from the X-Shooter 
spectra, in the following photometric systems: 
$UBVRI$ Bessel \citep{1990PASP..102.1181B}, 
$JHK_s$ 2MASS \citep{2006AJ....131.1163S}, 
$ugriz$ SDSS \citep{1990PASP..102.1181B} and 
again $JHK_s$ 2MASS \citep{2006AJ....131.1163S}, but converted to AB 
magnitudes.}
\begin{center}
\begin{small}
\begin{tabular}{@{}l@{}c@{}c@{}c@{}c@{}c@{}c@{}c@{}@{}c@{}c@{}c@{}c@{}c@{}c@{}c@{}}
\hline
\noalign{\smallskip}
\multicolumn{1}{c}{Galaxy} &
\multicolumn{1}{c}{$U$$-$$B$} &
\multicolumn{1}{c}{$B$$-$$V$} &
\multicolumn{1}{c}{$V$$-$$R$} &
\multicolumn{1}{c}{$R$$-$$I$} &
\multicolumn{1}{c}{$I$$-$$J$} &
\multicolumn{1}{c}{$J$$-$$H$} &
\multicolumn{1}{c}{$H$$-$$K_s$} &
\multicolumn{1}{c}{$u$$-$$g$} &
\multicolumn{1}{c}{$g$$-$$r$} &
\multicolumn{1}{c}{$r$$-$$i$} &
\multicolumn{1}{c}{$i$$-$$z$} &
\multicolumn{1}{c}{$z$$-$$J$} &
\multicolumn{1}{c}{$J$$-$$H$} &
\multicolumn{1}{c}{$H$$-$$K_s$} \\
\noalign{\smallskip}
\multicolumn{1}{c}{ID} &
\multicolumn{1}{c}{[mag]} &
\multicolumn{1}{c}{[mag]} &
\multicolumn{1}{c}{[mag]} &
\multicolumn{1}{c}{[mag]} &
\multicolumn{1}{c}{[mag]} &
\multicolumn{1}{c}{[mag]} &
\multicolumn{1}{c}{[mag]} &
\multicolumn{1}{c}{[mag]} &
\multicolumn{1}{c}{[mag]} &
\multicolumn{1}{c}{[mag]} &
\multicolumn{1}{c}{[mag]} &
\multicolumn{1}{c}{[mag]} &
\multicolumn{1}{c}{[mag]} &
\multicolumn{1}{c}{[mag]} \\
\noalign{\smallskip}
\hline
\noalign{\smallskip}  
NGC\,0584 &    0.53 & 1.09 & 0.58 & 0.65 & 1.18 & 0.73 &    0.40 & 1.90 & 0.90 & 0.43 & 0.36 & 0.40 & 0.25 & $-$0.07 \\
NGC\,0636 &    0.51 & 1.05 & 0.55 & 0.56 & 1.18 & 0.74 &    0.39 & 1.86 & 0.86 & 0.37 & 0.32 & 0.41 & 0.26 & $-$0.08 \\
NGC\,0897 &    0.38 & 1.15 & 0.56 & 0.52 & 1.30 & 0.78 &    0.42 & 1.72 & 0.92 & 0.34 & 0.29 & 0.54 & 0.30 & $-$0.05 \\
NGC\,1357 &    0.45 & 1.09 & 0.59 & 0.63 & 1.35 & 0.75 &    0.40 & 1.81 & 0.92 & 0.41 & 0.35 & 0.58 & 0.28 & $-$0.07 \\
NGC\,1425 &    0.36 & 0.94 & 0.49 & 0.53 & 1.18 & 0.68 &    0.04 & 1.68 & 0.71 & 0.35 & 0.30 & 0.41 & 0.20 & $-$0.44 \\
NGC\,1600 &    0.39 & 0.98 & 0.51 & 0.64 & 1.37 & 0.69 & $-$0.01 & 1.70 & 0.75 & 0.41 & 0.35 & 0.62 & 0.22 & $-$0.48 \\
NGC\,1700 &    0.53 & 1.07 & 0.53 & 0.60 & 1.11 & 0.71 &    0.03 & 1.91 & 0.83 & 0.39 & 0.36 & 0.31 & 0.24 & $-$0.45 \\
NGC\,2613 &    0.63 & 1.22 & 0.62 & 0.83 & 1.45 & 0.78 &    0.01 & 2.07 & 0.99 & 0.54 & 0.44 & 0.68 & 0.31 & $-$0.46 \\
NGC\,3115 &    0.57 & 1.11 & 0.44 & 0.56 & 1.34 & 0.75 &    0.23 & 1.96 & 0.82 & 0.18 & 0.46 & 0.57 & 0.28 & $-$0.24 \\
NGC\,3377 &    0.50 & 1.00 & 0.51 & 0.61 & 1.26 & 0.73 &    0.38 & 1.86 & 0.76 & 0.39 & 0.33 & 0.50 & 0.25 & $-$0.09 \\
NGC\,3379 &    0.56 & 1.02 & 0.49 & 0.62 & 1.24 & 0.70 &    0.00 & 1.93 & 0.74 & 0.40 & 0.34 & 0.47 & 0.22 & $-$0.47 \\
NGC\,3423 &    0.13 & 0.81 & 0.46 & 0.48 & 1.21 & 0.70 &    0.33 & ...  & 0.63 & 0.33 & 0.25 & 0.46 & 0.23 & $-$0.15 \\
NGC\,4415 &    0.25 & 0.85 & 0.46 & 0.48 & 1.19 & 0.68 & $-$0.05 & 1.47 & 0.64 & 0.33 & 0.25 & 0.43 & 0.20 & $-$0.52 \\
NGC\,7424 & $-$0.34 & 0.51 & 0.29 & 0.09 & 1.06 & 0.60 &    0.24 & 0.67 & 0.34 & 0.04 & 0.10 & 0.32 & 0.13 & $-$0.23 \\
\noalign{\smallskip}
\hline
\noalign{\medskip}
\end{tabular}
\end{small}
\label{tab:colors}
\end{center}
\end{table*}

\end{appendix}

\end{document}